\documentclass[12pt,preprint]{aastex}

\accepted{}
\journalid{}{}
\articleid{}{}

\shortauthors{Mohanty et al..}
\shorttitle{Measuring Masses of Substellar Objects}

\received{2003 March}
\begin{document}

\def\tross{${\tau}_{R}$ }
\def\hal{{H\alpha} }
\def\rc{R_C}
\def\ic{I_C}
\def\rcm{R_{Cm}}
\def\icm{I_{Cm}}
\def\jm{J_{m}}
\def\rco{R_{Co}}
\def\ico{I_{Co}}
\def\jo{J_{o}}
\def\av{A_V}
\def\ar{A_{R_C}}
\def\ai{A_{I_C}}
\def\exj{A_{J}}
\def\kr{k_{R_C}}
\def\ki{k_{I_C}}
\def\rad{\mathcal{R}}
\def\dist{\mathcal{D}}
\def\mass{\mathcal{M}}
\def\lum{{\mathcal{L}}_{bol}}
\def\logten{log_{10}}
\def\na{\ion{Na}{1}\ }
\def\pot{\ion{K}{1}\ }

\def\kms{km s$^{-1}$\ }
\def\kmsp{km s$^{-1}$ pix$^{-1}$\ }
\def\cms{cm s$^{-1}$\ }
\def\cmc{cm$^{-3}$\ }
\def\cmss{cm$^{2}$ s$^{-1}$\ }
\def\cmcs{cm$^{3}$ s$^{-1}$\ }

\def\msun{M$_\odot$\ }
\def\rsun{R$_\odot$\ }
\def\lsun{L$_\odot$\ }
\def\mj{M$_J$\ }

\def\teff{T$_{e\! f\! f}$~}
\def\gv{{\it g}~}
\def\vsini{{\it v}~sin{\it i}~}
\def\vrad{v$_{\it rad}$~}
\def\lbol{L_{\it bol}}
\def\lhal{L_{H\alpha}}
\def\fhal{F_{H\alpha}}
\def\fbol{F_{\it bol}}
\def\lx{L_{X} }
\def\eqwhal{EW_{H\alpha}}
\def\alom{{\alpha}{\Omega} }
\def\ross{R_{0} }
\def\cots{{\tau}_{c} }
\def\fchal{{\mathcal{F}}_{c\hal} }
\def\h2o{H$_2$O}

\title{Measuring Fundamental Parameters of Substellar Objects. II: Masses and Radii}

\author{Subhanjoy Mohanty\altaffilmark{1},  Ray Jayawardhana\altaffilmark{2}, Gibor Basri\altaffilmark{3}}

\altaffiltext{1}{Harvard-Smithsonian Center for Astrophysics, Cambridge, MA 02138.  smohanty@cfa.harvard.edu}
\altaffiltext{2}{Astronomy Department, University of Michigan, Ann Arbor, MI 48109.  rayjay@umich.edu}
\altaffiltext{3}{Astronomy Department, University of California, Berkeley, CA 94720.  basri@soleil.berkeley.edu}

\begin{abstract} 
We present mass and radius derivations for a sample of very young, mid- to late M, low-mass stellar and substellar objects in Upper Scorpius and Taurus.  In a previous paper, we determined effective temperatures and surface gravities for these targets, from an analysis of their high-resolution optical spectra and comparisons to the latest synthetic spectra.  We now derive extinctions, radii, masses and luminosities by combining our previous results with observed photometry, surface fluxes from the synthetic spectra and the known cluster distances.  These are the first mass and radius estimates for young, very low mass bodies that are {\it independent} of theoretical evolutionary models (though our estimates do depend on spectral modeling).  We find that for most of our sample, our derived mass-radius and mass-luminosity relationships are in very good agreement with the theoretical predictions.  However, our results diverge from the evolutionary model values for the coolest, lowest-mass targets: our inferred radii and luminosities are significantly larger than predicted for these objects at the likely cluster ages, causing them to appear much younger than expected.  We suggest that uncertainties in the evolutionary models - e.g., in the choice of initial conditions and/or treatment of interior convection - may be responsible for this discrepancy.  Finally, two of our late-M objects (USco 128 and 130) appear to have masses close to the deuterium-fusion boundary ($\sim$9--14 Jupiters, within a factor of 2).  This conclusion is primarily a consequence of their considerable faintness compared to other targets with similar extinction, spectral type, and temperature (difference of $\sim$ 1 mag).  Our result suggests that the faintest young late-M or cooler objects may be significantly lower in mass than current theoretical tracks indicate.  
\end{abstract}


\section{Introduction}
The last few years have witnessed a dramatic swelling in the ranks of objects at the bottom of the Main Sequence, and in the substellar regime beyond.  Hundreds of ultra-low-mass stars and brown dwarfs have been uncovered, both in the field and in star-forming regions.  Studies of young clusters even suggest the presence of isolated planetary mass objects \citep{Zapa00, Lucas00}.  The existence and properties of all these low-mass bodies have profound implications for a host of issues, ranging from the dominant mechanisms for star and planet formation \citep{Boss01, Reipurth01, Bate02, Padoan02}, to the birthline and early evolution of low-mass objects \citep{Hartmann03}, to the shape of the initial mass function \citep{Briceno02}.  A reliable determination of mass is intrinsic to the ultimate resolution of these questions.

Presently, masses (and ages) are most widely inferred by comparing observables such as temperature and luminosity to the predictions of theoretical evolutionary tracks.  However, these models remain largely unverified for very low masses.  The simplest test is to derive dynamical masses for the components of binary (or higher-order) systems with known orbital parameters, and compare them to the theoretical values derived from other, directly observed quantities (e.g., $\lbol$ and \teff).  Unfortunately, this is impeded for very low-mass stars and substellar objects by the current paucity of suitable multiple systems.  In most known cases, one can either deduce dynamical masses but not theoretical ones (because the components are not directly detected), or vice versa (because the orbital parameters remain indeterminate).  The one exception is HD 209458, in which both are available \citep{Charbonneau02}.  The comparison of theory to observations in this case does reveal some large uncertainties in the former, and underlines the usefulness of such tests (Baraffe et al. 2003; Burrows, Sudarsky \& Hubbard 2003).  However, it is not very illuminating as a general evaluation of the models: the proximity between the planetary companion and the star in this instance engenders special insolation effects, precluding an extension of the results to free-floating brown dwarfs and planetary mass objects (or to planets with larger orbital radii).  
 
The situation is likely to improve in the near future, at least in the field - several promising systems with directly detected, probably substellar components have now come to light; dynamical masses should be obtained fairly soon \citep{Close02, Potter02, Lane01}, allowing checks on the theoretical models for field brown dwarfs.  In young clusters and star-forming regions, however, no suitable systems have emerged yet.  This is especially troubling since even the identification of objects as substellar currently depends, at these early ages, on the theoretical tracks (empirical tests of substellarity that depend on Lithium detection or minimum Main Sequence temperature are largely inapplicable to very young objects).  Moreover, the low-mass tracks are most uncertain precisely at such early times \citep{Baraffe02}, so testing them for young objects is particularly crucial.  

To address this issue, we have developed a technique for calculating masses for young cluster objects from surface gravity measurements, independent of theoretical evolutionary models.  The essential idea is simple: derive surface gravity and effective temperature (\teff) by comparing the observed spectrum to the latest synthetic ones, then derive radius (and extinction) by combining the observed photometry and known cluster distance with the surface fluxes predicted by the synthetic spectra (for the inferred \teff and gravity), and finally derive mass by combining radius and gravity.  The sticking point, of course, is the derivation of sufficiently accurate surface gravities from the spectra, which has long been one of the major goals in the study of ultra-low mass objects.  However, we have shown in a previous paper (Mohanty et al. 2003a; henceforth Paper I) that the current generation of highly detailed synthetic spectra is equal to the task.  Employing these, we have derived gravities to within $\pm$ 0.25 dex (and \teff to within $\pm$ 50K) in a sample of very low-mass objects in the Upper Scorpius and Taurus clusters (Paper I).  We now derive masses and radii for these, using our Paper I results together with photometry and distance estimates.  We will show that our analysis allows mass to be determined to within a factor of $\sim$ 2, and radius to within $\sim$ 30\%.  These errors are much larger than those associated, for example, with dynamical mass and radius measurements in eclipsing binaries.  Nevertheless, we will demonstrate that they are sufficient for first order tests of the theoretical evolutionary tracks.  

Though our analysis is independent of the evolutionary models (and thus serves as a check on the latter), it is clearly dependent on the validity of the synthetic spectra we use.  The accuracy of these were discussed in Paper I, and will be addressed further in this work.  However, we point out that our derivation of physical parameters using spectral synthesis alone does not constitute a great leap of faith, any more than employing evolutionary models for this purpose does.  There are two reasons for this.  First, the $P$-$T$ structure of the deep atmosphere (which forms the inner boundary of the spectral calculations) acts as the outer boundary condition of the interior calculations; i.e., the evolutionary models are anchored with the same (deep) atmospheric modeling as the synthetic spectra.  Second, in order to compare observations to the evolutionary predictions, synthetic spectra are crucial: either to convert an observed spectral type to \teff (when placing objects on a theoretical \teff-luminosity H-R diagram), or to convert predicted effective temperatures and luminosities to photometric colors and magnitudes (when placing objects on a theoretical color-magnitude diagram).  We have only taken the dependence on synthetic spectra a step further, by using them to derive surface gravities as well; the accuracy of our gravities is discussed at length in Paper I.  The advantage of this route lies in our avoiding (and thereby testing) what are perhaps the greatest uncertainties in the theoretical models for young objects: the initial conditions and still-relevant effects of accretion and collapse during the formation stage. The main disadvantage of eschewing the evolutionary models is that we cannot independently estimate ages.  This is compensated for by our ability to independently estimate the mass, and thus provide a check on the evolutionary model predictions for a group of objects that belong to the same cluster, and are thus likely to be (nearly) coeval. 

Finally, we elucidate the system of nomenclature we have adopted here for very low-mass objects. There are considerable differences within the community at present regarding this issue: naming conventions based on fusion (or equivalently, mass), origins and location (isolated or in orbit around a star) have all been proposed.  Controversies arise because the different definitions do not yield the same grouping of objects; which of these systems is finally adopted is a matter for future arbitration.  However, given that a consensus is currently lacking, and that our primary concern in this paper is mass, we adopt a fusion-based convention (since mass is most directly associated with the presence and type of fusion).  The term `brown dwarf' refers to all objects which never derive 100\% of their luminosity from hydrogen burning (unlike stars), but which are nevertheless above the deuterium-burning mass limit.  Thus brown dwarfs are objects in the range $\sim$ 0.012--0.080 \msun (12--80 \mj).  We contract the term `planetary mass object' to the less cumbersome `planemo', and use it to refer to all objects below the deuterium-burning limit (i.e., mass $\lesssim$12 \mj), regardless of whether they are free-floating or in orbit around a star.  When a distinction is required between the two cases (e.g., when referring specifically to the recent isolated planemo candidates), it will be made explicitly.  Since neither planemos nor brown dwarfs undergo stable hydrogen fusion (i.e., reach the Main Sequence), both are included under the rubric of `substellar objects'.  Since both stars and brown dwarfs undergo at least some fusion, they will collectively be termed `fusors'; in this context, planemos are non-fusors.  A broader discussion of these terms and nomenclature issues is presented in Basri 2003.  

\section{Overview of Sample and Observations}
A detailed account of our sample selection, observations and data reduction methods, and evidence for cluster membership has been presented in Paper I (also see Jayawardhana, Mohanty \& Basri 2002, 2003).  We only cite the salient points here.  Our sample consists of 11 low-mass Pre-Main Sequence (PMS) objects in Upper Scorpius, ranging in spectral class from M5 to M7.5, and 2 low-mass PMS objects in Taurus - GG Tau Ba ($\sim$ M5.5-6) and Bb (M7.5), which form a close binary within the quadruple system GG Tau.  High-resolution optical spectra were obtained for all targets using HIRES on Keck I.  The cluster membership and PMS status of the Upper Sco targets were verified using four criteria:  presence of Lithium absorption (LiI 6708\AA), radial velocity consistent with that of known Upper Sco members, presence of strong H$\alpha$ emission analogous to that in field dMe stars, and neutral alkali (\na and \pot) line strengths intermediate between those in field dwarfs and giants of similar spectral type.  11 out of an original list of of 14 Upper Sco candidates met all four criteria, and are included in our final sample.  GG Tau Ba and Bb are previously confirmed PMS members of Taurus; we did not undertake detailed membership tests in their case.  The present work also requires photometric and distance measurements.  For the Upper Sco targets, we adopt NIR photometry from 2MASS, and optical photometry from Ardila, Mart\'{i}n \& Basri (2000; hereafter AMB00, from whose initial photometric survey of Upper Sco our sample is culled).  For GG Tau Ba and Bb, we use the optical and NIR photometry cited by White et al. (1999; hereafter WGRS99).  Average distances to the Upper Sco and Taurus regions are taken from Preibisch et al. 2002 and WGRS99 respectively.

\section{Extinction, Radius and Mass Analysis}
Our method of analysis may be summarized thus.  We derive extinctions by comparing the observed optical colors ($\rc$ - $\ic$) to those predicted by synthetic spectra for the appropriate \teff and gravity.  The radii are then inferred via two slightly different methods.  In the first, they are derived by combining the extinctions with the observed $\ic$-band flux, the predicted $\ic$-band flux at the stellar surface, and the known distance to the cluster.  In the second, the same procedure is followed, but now using $J$-band fluxes instead of $\ic$-band ones.  Finally, masses are acquired from the radii and the previously derived surface gravities, and luminosities from the radii and previously inferred \teff.  Thus, for each object, we have one estimate of extinction from $\rc-\ic$, but two each of radius, mass and luminosity, one from using $\ic$ to determine radius and one from using $J$.  We elaborate on this method in \S 3.1 below, and quantify our expected stochastic errors in \S 3.2; the important sources of potential systematic errors are discussed in \S 3.3.  Finally, the rationale for employing $\rc-\ic$ for the extinction calculations, and $\ic$ or $J$ for the radius ones, is also discussed in \S 3.3.    

\subsection{Method of Analysis}
In Paper I, we derived \teff and log \gv for our sample by comparing our high-resolution optical spectra to the latest synthetic spectra by Allard \& Hauschildt.  The latter have been discussed in detail in Paper I.  For our purposes here, it suffices to note that the model spectra have been constructed using plane-parallel atmospheres.  As such, they are independent of the stellar radius or mass; they depend only on \teff and gravity, and predict the emergent flux per unit wavelength at the stellar surface.  Convolving the latter with appropriate bandpasses gives the predicted apparent magnitude {\it at the stellar surface} in any desired photometric band, for a specified \teff and gravity.  This convolution has been carried out for the models, for standard optical and near-infrared passbands\footnote{Available at ftp.ens-lyon.fr/pub/users/CRAL/fallard/AMES-Cond-2002/colmag.AMES-Cond.opt6 and .../colmag.AMES-Cond-2002.ukirt}.  For our sample, we also possess observed optical photometry in the Cousins $\rc$~ and $\ic$~ bands, as well as NIR photometry in the $J$ band.  Temperatures, gravities and photometric measurements are listed in Table 1.  From these, we calculate extinctions, radii, masses and luminosities as follows.

$$ Extinction:\qquad\qquad\qquad \av \,=\, \frac{(\rco - \ico) \,-\, (\rcm - \icm)}{\kr-\ki} \qquad\qquad\qquad\qquad\qquad\eqno[1] $$
$${\rm where}\,\,\,\,\,\,\,\,\, \kr \,\equiv\, \frac{\ar}{\av} \,=\,0.81 \,\, , \,\, \ki \,\equiv\, \frac{\ai}{\av} \,=\, 0.60 \qquad \,\eqno[2] $$ 
Here, $\rco$~ and $\ico$~ are the observed magnitudes for a given object.  $\rcm$~ and $\icm$ are the synthetic magnitudes at the stellar surface, for the best-fit synthetic spectrum to that object found from our previous \teff and gravity analysis.  $\av$~, $\ar$~ and $\ai$ are the extinctions at $V$, $\rc$~ and $\ic$~ respectively.  The extinction ratios ($\kr$~, $\ki$~) given in equation [2] are taken from Schlegel, Finkbeiner \& Davis 1998; they are appropriate for CCD $\rc$ and $\ic$ filters, and a normal extinction law ($R_V$ $\equiv$ $\av/E(B-V)$ $\approx$ 3.1).  They imply that $\av$=4.76$E(\rc-\ic)$, in agreement with the relation used by other authors for Upper Sco (e.g., Preibisch et al. 2002).  Indeed, $R_V$ $\approx$ 3.1 appears to be valid even for the heavily reddened Taurus-Auriga region \citep{Kenyon94}.  We therefore assume that a normal extinction law holds for both our Upper Sco sample (following Preibisch et al. 2002), as well as for GG Tau B (in agreement with WGRS99).  Note that in the most commonly used procedure for determining extinction, the observed spectrum is first compared to various stellar templates over narrow spectral regions (over which the differential reddening is negligible) to determine the spectral type; the extinction is then derived by comparing the observed colors to unreddened ones at that spectral type.  Our method is completely analogous, except that we use synthetic spectra as templates (instead of observed stellar spectra of similar spectral type).  Our extinctions are listed in Table 1.  

$$ Radius:\qquad\qquad\qquad \logten\left[\rad\right] \,=\, \logten\left[\dist\right] \,+\, \left[\frac{\icm \,-\, (\ico \,-\, \ai)}{5}\right]  \qquad\qquad\qquad\eqno[3a] $$
or alternatively,

$$\quad\qquad\qquad\qquad\qquad\qquad \logten\left[\rad\right] \,=\, \logten\left[\dist\right] \,+\, \left[\frac{\jm \,-\, (\jo \,-\, \exj)}{5}\right]  \qquad\qquad\qquad\eqno[3b] $$

Here, $\rad$~ is the stellar radius and $\dist$~ the distance to the star (both in parsecs).  We adopt $\dist$~ = 140pc for GG Tau and 145 pc for our entire Upper Sco sample; these are, respectively, the mean distances to the Taurus star-forming region and to the Upper Sco association.  $\ai$~ in eqn. [3a] is calculated from the $\av$ (eqn. [1]), using eqn. [2].  ($\ico-\ai$) is then the $\ic$ magnitude that would be observed on Earth in the absence of extinction; we will often refer to this extinction-corrected value as the `intrinsic $\ic$-band flux'.  Similarly, $\jm$, $\jo$ and $\exj$ in eqn. [3b] are the synthetic $J$ magnitude at the stellar surface, the observed $J$ magnitude, and the extinction in the $J$-band respectively; $\exj$ is derived from the $\av$ (eqn. [1]) using $\exj / \av$  = 0.28.  ($\jo-\exj$) is then the $J$ magnitude expected on Earth in the absence of extinction; as in the $\ic$-band case, we will refer to this extinction-corrected value as the `intrinsic $J$-band flux'.  The logic behind eqns. [3] is simple: for an unresolved source, in any photometric band, the translation between flux at the stellar surface and the flux finally observed depends upon radius, distance and extinction; any one of these quantities (in our case, radius) can be calculated if all the others are known.  We also point out again that, regardless of whether radius is inferred through $\ic$ fluxes or $J$ (eqn. [3a] or [3b]), the extinction used in its derivation is always from $\rc-\ic$ color (i.e., though either $\ai$ or $\exj$ is used in the radius equations, these are not independently measured quantities but simply calculated from the $\av$ found through eqn. [1]; thus $\ai$ and $\exj$ depend only on $\rc-\ic$).

{\it Mass}:  Mass ($\mass$) is inferred from the radius and previously derived surface gravity, through Newton's law of gravitation.   

{\it Luminosity}:  Luminosity ($\lum$) is inferred from the radius and previously derived \teff.  

Clearly, the two radius estimates, one from $\ic$ and the other from $J$ (eqns. [3a] and [3b]), yield two values for mass and luminosity.  The rationale for making independent estimates based on $\ic$ and $J$ comes from potential sytematics in the synthetic photometry; this is discussed in detail in \S 3.3 and Appendix A.  Both $\ic$ and $J$-based values are listed in Table 2.  In reality, it turns out that the two sets of estimates are in close agreement for most objects: using $\ic$ or $J$ makes no significant difference.  The only exceptions are our faintest targets, in which uncertainties in the {\it observed} optical photometry appear responsible for a divergence between $\ic$ and $J$.  This is addressed in \S3.3 and Appendix B; as we argue there, the $J$-based values should be more accurate for these objects.  Consequently, $J$ estimates are good for all our targets; we therefore plot only the $J$-based values in Figs. 1--4.

\subsection{Stochastic Errors}
For PMS gravities (log \gv $\sim$ 3.0--4.0), $\rc-\ic$ color in the models changes by $\sim$ 0.1 mag for a 100K change in \teff, and by $\lesssim$ 0.1 mag for a 1 dex change in gravity.  Since our internal errors in derived \teff and gravity are $\sim$ $\pm$50K and $\pm$0.25 dex (Paper I), the corresponding uncertainties in model $\rc-\ic$ color are $\sim$ $\pm$0.05 mag and $\pm$0.025 mag, from \teff and log \gv uncertainties respectively.  Finally, for Upper Sco, the measurement errors in observed $\rc-\ic$ (see AMB00) are at most of order $\pm$0.12 mag.  Adding all these in quadrature gives a total error of $\sim$ $\pm$0.13 mag; we adopt $\pm$0.14 mag (equivalent to adopting 0.1 mag errors both in observed color, and in intrinsic model color due to combined \teff and gravity uncertainties).  This leads to an error of $\pm$0.7 mag in $\av$ (see eqns. [1] and [2]), which translates to an uncertainty of $\pm$ 0.42 mag in $\ai$ and $\pm$ 0.20 mag in $\exj$.  

We also have errors of $\sim$ $\pm$ 0.15 mag in the model $\ic$, due to combined \teff and gravity uncertainties; we adopt 0.2 mag.  The errors in observed $\ic$ in Upper Sco (AMB00) are $\lesssim$ $\pm$0.1 mag; we adopt 0.1 mag.  Finally, combining Hipparcos data with an analysis of secular parallaxes constrains the variation about the mean distance to Upper Sco to $\lesssim$ 20pc, assuming a spherical geometry for the association \citep{Preb02}; this translates to $\lesssim$ $\pm$0.06 dex in $\logten\left[\dist\right]$.  Collecting all these in quadrature, and including the factor of 5 in the denominator of eqn. [3a], gives a final error of $\pm$0.11 dex in $\logten\left[\rad\right]$, i.e., $\pm$30\% uncertainty in $\ic$-based USco radii.  When $J$ fluxes are employed instead, the internal errors are smaller: while errors in synthetic photometry due to our \teff and log \gv uncertainties are similar in $J$ and $\ic$, extinction errors contribute less in $J$ than in $\ic$, and the errors in observed photometry are also lower in $J$ ($\lesssim$ $\pm$0.03 mag from 2MASS).  Consequently, our internal errors in $\logten\left[\rad\right]$ are then $\pm$0.08 dex, i.e., $\pm$20\% uncertainty in $J$-based USco radii.  
   
Mass is proportional to gravity, and to radius squared.  Our uncertainty in log \gv is $\pm$ 0.25 dex.  Combining this with our error in radius gives a final error in $\logten\left[\mass\right]$ of $\pm$0.33 or $\pm$0.30 dex, depending on whether radius is calculated from $\ic$ or $J$.  Our USco masses are thus precise to within a factor of $\sim$2 (2.1 for $\ic$, 2.0 for $J$).  Finally, our $\pm$50 K \teff error contributes negligibly to the luminosity uncertainty; the latter depends entirely on the error in radius.  Our errors in $\logten\left[\lum\right]$ are therefore $\sim$ $\pm$0.22 and $\pm$0.16 dex, for radius based on $\ic$ and $J$ respectively; i.e., our $\lum$ are uncertain by $\sim$ 55\% (65 \% for $\ic$, 45\% for $J$).  Finally, for all quantities, our errors for GG Tau Ba and Bb are are very similar to those for USco: the uncertainties in their observed photometry, from WGRS99, are comparable to those in USco, and all other errors in our analysis are the same for both clusters.  

As noted above, our $J$-based values have slightly smaller internal errors than the $\ic$ ones.  However, to err on the side of caution, we henceforth adopt the higher, $\ic$-based errors for the $J$ estimates as well.  Moreover, for both $\ic$ and $J$, the fact that our mass and luminosity estimates are dependent on our radius measurements has an important consequence for the plots shown in \S 4.  When plotting mass or luminosity against radius (or against each other), the X- and Y-axis errors are not independent in our analysis, but coupled.  Consequently, the (1$\sigma$) errors are illustrated with ellipses, rather than perpendicular error bars.  
  
Finally, we point out that even with a \teff uncertainty of 100K (twice our error in Paper I), we would still obtain radii to within $\sim$ 35\%, masses to within a factor of $\sim$ 2.3, and luminosities to within $\sim$ 70\%.  These precisions are not much worse than obtained above with a \teff uncertainty of 50K (using the same $\pm$0.25 dex error in gravity in both cases).  Our results are thus quite robust even with \teff errors comparable to those in analyses of low-resolution spectra (e.g., Leggett et al. 2000).  Our constraints on mass and radius, substantially better than previously reported from spectral analysis, arise mainly from our well-constrained gravity ($\pm$0.25 dex).  As demonstrated in Paper I, the synthetic spectra we use allow this level of precision in log \gv, even if \teff were uncertain by 100K instead of 50K.

Our foregoing analysis is concerned with internal stochastic errors, i.e., the {\it precision} of our measurements.  This does not address however, the absolute accuracy, or {\it veracity}, of our values.  To assess the latter, we now briefly discuss possible systematic offsets in our analysis.  In fact, as we will show, it is the potential for such offsets in the synthetic spectra that dictates the specific color and photometric bands we have chosen to derive $\av$ and radii.  

\subsection{Systematic Errors}
Our sources of systematic error fall into three categories: {\it (1)} offsets in the synthetic photometry, {\it (2)} systematic errors in the observed photometry, {\it (3)} systematic errors in our adopted \teff and gravity, and {\it (4)} real physical phenomena, most pertinently cool spots and binarity.  We summarize here the effects of each; details are presented in the Appendices.  

{\it Synthetic Photometry:} All the parameters in this paper are calculated by comparing the observed photometry and colors of our PMS M-type sample to those predicted by the synthetic spectra.  We test the validity of the latter models by comparing them to field M dwarfs.  Our conclusion is that, due to possible offsets in the model photometry, extinction ($\av$) is best calculated by using $\rc-\ic$ colors (eqn.[1]), while radius (and hence mass and luminosity) is most accurately derived from $J$-band fluxes (eqn.[3b], where the $\exj$ is directly computed from the $\av$ implied by eqn.[1]).  At the very least, the two radius estimates, from $\ic$ and $J$ respectively (eqns. [3a and b]), should bracket the true value very well.  In reality, we find that the radii, masses and luminosities derived from both $\ic$ and $J$ agree very well in the majority of cases (Table 2; $\lesssim$ 15\% difference in $\rad$, $\Rightarrow$ $\lesssim$ 30\% disparity in $\mass$ and $\lum$).  The only exceptions are objects in which the observed optical photometry appears to be incorrect, as we discuss next; in these, the $J$-based parameters are more accurate.  In conclusion, systematic offsets in synthetic photometry have very little effect on our radii, masses and luminosities.  The full analysis is presented in {\it Appendix A}.  

{\it Observed Photometry:} In our five faintest objects (USco 100, 109, 112, 128 and 130), the $J$-based radii are $\sim$ 20--40\% larger than the $\ic$ ones.  We find that this is due to anomalies in their observed fluxes and colors.  Comparisons using 2MASS $H$ and $K$ photometry, which are only marginally affected by extinction effects in these objects, reveals that our $J$-band calculations are likely to be quite accurate, while AMB00's reported $\rc$ and $\ic$ fluxes (which we use) appear to be uncertain for these five targets (as is quite plausible given their faintness in the optical, close to AMB00's completeness limits).  We therefore adopt the $J$-based radii, masses and luminosities for these objects.  A detailed analysis can be found in {\it Appendix B}.  

{\it \teff and Gravity:} Systematic errors in our derived \teff, from spectral analysis, will result in attendant offsets in the synthetic photometry we adopt for any given object, and thus potentially skew our radius calculation.  However, we show that such a \teff offset, at the 100--200K level, will in reality hardly affect our derived radii, masses and luminosities: it changes both the derived $\av$ as well as the adopted surface flux in a given filter, but in opposite senses, so that the net effect is minimal (eg, only a $\lesssim$10\% change in radius for a 200K shift in \teff).  Gravity offsets, meanwhile, have a negligible effect on the photometry, so do not affect our radii.  Errors in log \gv will certainly influence our calculated mass directly; however, we present arguments (as we have in Paper I as well) that significant systematics in our adopted gravities (i.e., larger than our adopted $\pm$ 0.25 dex measurement uncertainty) are very unlikely.  A comprehensive analysis of these issues is preesented in {\it Appendix C}.        

{\it Cool Spots and Binarity:} We show that even large cool spots (50\% areal coverage, 500K cooler than surrounding photosphere) affect our extinction, mass and radius calculations negligibly, while they make us underestimate luminosity by $\sim$ 25\% (mainly due to the 200K lower \teff we infer in the presence of such spots; see Paper I).  Large {\it ultra}-cool spots, covering a significant fraction of the stellar surface and appearing completely dark in a given photometric band compared to the photosphere, can influence our results: with a 50\% coverage by a such a spot, we will underestimate radius by a factor of $\sqrt{2}$, and mass and luminosity by a factor of 2 each.  Such spots are implausible in any significant fraction of our sample; nevertheless, we do examine if they can be responsible for the very lowest masses we derive (\S 4.1.1).  Binarity, meanwhile, has an effect analogous to cool spots.  For equal mass binaries, we will {\it over}estimate radius by $\sqrt{2}$, and mass and luminosity by a factor of 2.  This effect may be responsible for the very highest masses we derive, which appear a little too high for the corresponding spectral types (\S 4.2).  However, since we do not see any double-lined spectroscopic binaries in our sample, only very close companions (i.e., single-line systems) can skew our results; binarity should not be a large effect for our sample.  A full analysis of cool spots and binarity is presented in {\it Appendices D} and $E$ respectively.   

\section{Results}
With these introductory remarks, we move on to a detailed analysis of our mass, radius, \teff and luminosity results.  In the course of this, we will repeatedly compare our derived quantities to the predictions of the theoretical evolutionary tracks of the Lyon group.  Specifically, in order to encompass a large range in mass, we have combined the tracks presented in Baraffe et al. 1998 (hereafter, BCAH98) and Chabrier et al. 2000 (hereafter, CBAH00), as detailed in Paper I; following the convention in Paper I, we refer to this merged set of tracks as the Lyon98/00 models.  These tracks are the ones most widely used in the literature to infer the properties of young, very low-mass stellar and substellar bodies;  it is therefore particularly useful to check their predictions against our independently derived values.

Our extinctions, radii, masses and luminosities are listed in Tables 1 and 2.  Besides our Upper Sco targets, we have also derived parameters for the well-studied GG Tau B system (Ba and Bb), previously analysed in some detail by WGRS99.  For GG Tau Ba, our extinction, mass and luminosity estimates are very similar to those of WGRS99.  For Bb, however, our $\av$ and luminosity are significantly higher, while our mass is $\sim$40\% lower (due to the low gravity we find in Paper I).  Moreover, our extinction for GG Tau Bb is quite different from that for Ba, at odds with the similar $\av$ derived for both by WGRS99 (though their $\av$ for Ba and Bb are not the same as their values for Aa and Ab, the other two components of the GG Tau system).  A full discussion of GG Tau, supporting evidence for our extinctions, and detailed comparison to the WGRS99 results is presented in Appendix F.  In the following sections, we concentrate on the parameters derived in this paper.  

Finally, with regard to radius, mass and luminosity, our discussion will largely be limited to the $J$-based values: as noted earlier (\S 3.3), $\ic$ and $J$ yield similar values anyway for most of our objects (Table 2), while in the handful of exceptions (our faintest targets), the estimates derived from $J$ are likely to be more accurate.  

\subsection{Mass and Radius}
In Fig. 1, we plot our derived radii versus mass.  Also shown are the Lyon98/00 tracks for various ages, from 1 to 10 Myr.  Three striking facts are immediately evident.

First, two of our Upper Sco targets appear to lie close to the planemo boundary of 12 \mj: $J$ fluxes imply a mass of about 9 \mj for USco 128 and 14 \mj for USco 130.  Our masses are uncertain within a factor of 2, so they may well be brown dwarfs; even so, their position near the bottom of the brown dwarf mass sequence seems secure.  These two are our lowest gravity Upper Sco targets, as well as the faintest (Table 1); their ultra-low masses result from a combination of these two factors (e.g., GG Tau Bb, which has a very similar gravity and \teff but is much brighter, has a significantly higher mass of $\sim$ 25--30 \mj).  We discuss USco 128 and 130 in more detail in \S 4.1.1.  

Second, for masses $\gtrsim$ 0.03 \msun, our mass-radius relationship agrees remarkably well with that predicted by the Lyon98/00 tracks at the expected ages of Taurus (1--1.5 Myr) and Upper Sco (3--5 Myr).  However, third, there is a significant discrepancy between our radii and the predicted ones for the lowest masses ($\lesssim$ 0.03 \msun), with our values being considerably higher.  GG Tau Bb has a radius twice that expected for a 1--1.5 Myr old, 30 \mj object.  Similarly, for their masses, USco 128, 130 and 104 all have radii almost twice that predicted for an age of 3--5 Myr, and $\sim$40--50\% larger than expected even at 1 Myr.  For USco 104, the offset might conceivably arise from our analysis uncertainties: its position is consistent with the 3--5 Myr tracks within our errors in mass and radius.  For GG Tau Bb and USco 128 and 130, though, our result is robust in spite of the estimated errors.  This result can also be stated in terms of age: given our radii, the Lyon98/00 models suggest that our lowest mass objects are much younger than higher mass ones in the same clusters.  

Note that objects that agree/disagree with the track predictions in the mass-radius plane are the same ones that are most consistent/inconsistent with the tracks in our \teff-gravity plot in Paper I (see Fig. 9 in Paper I).  This is not coincidental, since our masses depend on gravity; we address this shortly in our analysis of the radius discrepancy (\S 4.1.2).     

It is noteworthy here that, regardless of our deviation from the Lyon98/00 tracks at the lowest masses, our derived masses and radii agree with two broad theoretical predictions, which are largely independent of the particular evolutionary models used.  One is that PMS objects of any given mass contract with age.  The other is that within a coeval sample, lower masses should generally have smaller radii (with some scatter introduced by any age spread, as well as by the details of the contraction process).  Our results clearly exhibit both trends:  the Taurus objects GG Tau Ba and Bb, which should be younger than the Upper Sco sample, are indeed somewhat larger than similar mass bodies in Upper Sco, while our inferred radii within both clusters also distinctly decrease with decreasing mass.  This result is heartening, given the spread in \teff, gravities and $\av$ in our sample, and the possible errors in the calculation of each; it bolsters our confidence in the derived parameters.  

We now analyze in some detail the two major results of our mass/radius calculations: the surprisingly low masses derived for USco 128 and 130, and the radius discrepancy with the tracks at the low-mass end of our sample.  In \S 4.1.1, we delve into the underlying reasons for the apparent near-planemo status of the two Upper Sco objects, and show that our result is reasonably robust in the face of various possible sources of error.  In \S 4.1.2, we examine whether systematic errors discussed earlier (\S 3.3) can explain our larger-than-predicted radii for the lowest masses, and conclude that they cannot; real problems seem to exist in the theoretical evolutionary models at these masses.  We then show, in \S 4.2, that the evolutionary tracks may be problematic even at the higher masses, where our results are so far in apparent agreement with the tracks.  Finally, we discuss our conclusions in in \S 5.
  
\subsubsection{Planemos}
We find a mass of 9 \mj for USco 128 and 14 \mj for USco 130, using $J$ fluxes ($\ic$ fluxes predict even lower masses, of 6 and 7 \mj respectively (Table 2); since the latter are likely to be spurious for these two faint targets (\S 3.3), we focus on the higher, more accurate $J$ values).  At the estimated age of Upper Sco (3--5 Myr), the theoretical evolutionary tracks predict that masses $\lesssim$ 15 \mj have \teff $\lesssim$ 2300K; planemos (mass $\lesssim$ 12\mj) have \teff $\lesssim$ 2100K; and masses $\lesssim$ 10\mj have \teff $\lesssim$ 2000K.  Our derived temperature for both objects, though, is about 2600K (comparable to our values for other mid- to late M's in the sample).  Our mass estimates thus present a rather severe temperature discrepancy (300--600K) with the theoretical expectations.  Moreover, we find USco 128 and 130 to have the same \teff, but much lower mass, compared to GG Tau Bb (whose \teff {\it is} more consistent with the tracks for its mass); this is at odds with the steep decline in \teff with decreasing mass predicted by the Lyon models (\S 4.2).  Our masses for USco 128 and 130 thus deserve greater scrutiny.

USco 128 and 130 have the lowest gravities in our Upper Sco sample, with log \gv $\approx$ 3.25 dex.  However, this is not the main reason behind their very low inferred masses.  This is best illustrated by comparing them to GG Tau Bb.  All three have similar spectral types ($\sim$M7--7.5), and our \teff from spectral analysis are correspondingly nearly identical.  Moreover, their gravities are all about the same; if anything, our gravity for GG Tau Bb (log gv $\approx$ 3.125) is slightly {\it lower} than in the other two (Table 1).  Despite these close similarities, our mass for GG Tau Bb is $\sim$ 30 \mj: a value which is not particularly remarkable, and more importantly, is 2--3 times larger than our estimates for USco 130 and 128 respectively.  Low gravity alone, therefore, does not account for the ultra-low masses we find in the latter two targets.  Instead, the primary reason is their considerable faintness compared to GG Tau Bb.  After correcting for extinction, we find USco 130 to be 1 mag fainter than Bb in $J$, while USco 128 is 1.5 mag fainter.  Since our distance and \teff for all three are nearly identical, we are led to ascribe this faintness to a reduction in surface area; consequently, our radii for USco 130 and 128 are much smaller (by a factor of 1.6 and 2 respectively) than for GG Tau Bb.  Since our gravities for the three are nearly the same, our masses derived using $\mass$ $\propto$ \gv$\rad$$^2$ are correspondingly far lower in the USco objects than in Bb.  Specifically, compared to Bb, $\rad$$^2$ in USco 130 and 128 is lower by a factor of 2.5--4, while their gravities are only a factor of 1.3 (0.125 dex) higher; hence they come out to be 2--3 times less massive than Bb. 
   
Of course, even if USco 128 and 130 were the same mass as GG Tau Bb, we would expect their radii to be smaller through contraction, since they are older.  However, the evolutionary models predict that this is not a very large effect, in the mass range of interest here.  Specifically, WGRS99 have found GG Tau Bb to have \teff $\sim$ 2800K and mass $\sim$ 40 \mj, by assuming that the theoretical evolutionary tracks are accurate.  Without making any such assumption, we have derived \teff $\sim$ 2600K and mass $\sim$ 30 \mj.  It appears safe to assume, therefore, that GG Tau Bb lies somewhere in this range (though we prefer our values).  From Fig. 1, we see that objects in this mass and \teff interval are expected to shrink in radius by at most $\sim$ 25\% (0.1 dex), in going from an age of $\sim$ 1 to 5 Myr.  The same conclusion can be drawn from Fig. 9 in Paper I, which shows that the predicted gravity increases from 1--5 Myr by at most 0.25 dex, and usually by much less, for these masses/temperatures.  Our derived log \gv difference of 0.125 dex between GG Tau Bb and the two USco objects is thus fully consistent with this expectation (though our {\it absolute} values of log \gv for these three objects diverge substantially from the model predictions, which is one of the main results of Paper I); as we have seen, planemo masses are derived in spite of this (small) increase in gravity.  Even the maximum, $\sim$25\% contraction predicted by the model tracks is insufficient to explain the factor of 1.6--2 decrease in radius in going from GG Tau Bb to USco 130 and 128, implied by their observed difference in brightness.  

It may be suggested that the difference in brightness between GG Tau Bb on the one hand, and USco 128 and 130 on the other, is not intrinsic, but is an artifact of inaccuracies in our derived $\av$: the extinctions we use for correcting the observed $J$ fluxes in USco 128 and 130 are derived from AMB00, and we have shown in Appendix B that these $\av$ values are probably somewhat incorrect, due to errors in AMB00's $\rc$ and $\ic$ photometry in the faintest objects.  However, we have also argued in Appendix B that any resulting $\av$ offsets are unlikely to significantly alter the intrinsic $J$ fluxes we have derived.  In particular, we have shown that comparisons using $H$ and $K$ photometry, which are even less affected by extinction than $J$, strongly support the claim that USco 130 and 128 are indeed intrinsically 1 and 1.5 mag fainter than GG Tau Bb respectively.  The $\av$ we have derived modify the observed $J$, $H$ and $K$ magnitudes only marginally; even if we assume $\av$=0 for all three objects (since our extinctions may be overestimated; Appendix A), USco 130 remains 1 mag fainter than GG Tau Bb in the NIR, while USco 128 is still $\sim$ 1.3 mag fainter (implying a mass of 11 \mj, quite similar to the 9 \mj derived using our $\av$) (Appendix B).  Realistic differences in $\av$ cannot account for this; USco 128 and 130 do appear intrinsically faint.  

Similarly, it is unlikely that large ultra-cool spots make the two USco objects anomalously faint: such spots would have to cover 60--75\% of the surface, and remain dark compared to the photosphere all the way out to $K$.  Such extremely large and cool spots are unlikely in any one object in our relatively small sample, let alone in two: indeed, the cool spot in this case would constitute the real photosphere, with the `true' photosphere reduced to a `hot spot'.  Distance uncertainties alone also seem incapable of producing the underluminosity of USco 128 and 130 compared to Bb.  Distances to both Upper Sco (145 pc) and Taurus (140 pc) are uncertain by $\sim$ 20pc; USco 130 and 128 thus remain underluminous by a factor of 1.3--2 even if we simultaneously assume that GG Tau is closer by 20 pc, and USco 128 and 130 farther by the same amount.  Finally, one may invoke the exotic scenario of close-to edge-on disks around USco 128 and 130.  Such orientations however, are very rare, and it is unlikely that two our targets are affected by them.  Moreover, neither of these objects shows strong signatures of ongoing accretion (JMB02), and there is no evidence for any excess $K$-$L'$ disk emission at least around USco 130 (USco 128 shows a moderate $K$-$L'$ excess; Jayawardhana et al. 2003).  Together, these facts make it rather improbable that nearly edge-on disks cause the underluminosity of both USco 128 and 130.

It is true that a combination of offsets in gravity, distance and photometry can lead to USco 128 and 130 having significantly higher masses, more consistent with that of GG Tau Bb, even if individually these factors are insufficient for this purpose.  This is illustrated in Fig. 1, where our upper limits in mass for the two USco objects are seen to be compatible (or nearly so) with Bb.  However, factors of 2--3 underestimations in mass are not apparent in any of our other targets; if anything, a few of them appear somewhat too massive (perhaps due to binarity).  Thus, while we cannot completely rule out USco 128 and 130 having masses close to that of GG Tau Bb, we believe that our much lower estimates, close to the planemo boundary, are fairly robust.  Our conclusion certainly needs to be thoroughly checked through further observations; the (distant) possibility of edge-on disks should also be pursued.  

\subsubsection{Analysis of Radius Discrepancy}
The second important result from our mass-radius analysis is that our coolest, lowest mass objects seem to have larger radii than predicted for their masses, at the expected cluster ages.  In \S 3.3.3, we have outlined various systematics that may influence our results. We now examine whether any of these can lead to the radius discrepancy, and find that they cannot.

Notice first that, apart from uncertainties in gravity (which we return to shortly), all the other systematics  - errors in synthetic and observed photometry, \teff offsets, binarity and cool spots - affect mass only through their influence on radius (since we use Newton's law to derive mass from radius and gravity).  That is, at a given gravity (inferred from spectral analysis), all these systematics cause our targets to move along locii on which $\mass$ $\propto$ $\rad$$^2$.  

Next, notice that at a given age, the evolutionary models predict approximately the same gravity for masses ranging from planemos to well into the stellar domain (Fig. 9, Paper I).  The only deviations from this are in the stellar/brown dwarf regime at the earliest ages (1--2 Myr), where assumptions about initial conditions related to deuterium-burning produce some larger variations in gravity (Paper I), and in very low-mass planemos, where the early onset of degeneracy makes log \gv decrease with mass (due to nearly constant radius).  To restate: while the gravity of a given mass increases with age as it contracts, gravity is predicted to generally remain nearly constant with mass (or equivalently, \teff) at a given age (for the range of masses, \teff and ages of interest here).  The mass-radius tracks simply reflect this near constancy in gravity: i.e., a track at a specified age closely traces a locus of $\mass$ $\propto$ $\rad$$^2$, where the constant of proportionality is the gravity predicted for that age (this is why the tracks are nearly straight lines in the logarithmic mass-radius plot shown in Fig. 1, except at the youngest ages and at low planemo masses).
  
Taken together, these considerations have two implications.  First, the aforementioned systematics, which directly influence radius, only cause our targets to slide {\it parallel} to the theoretical tracks in the mass-radius plane.  Thus, invoking these systematics cannot improve (or detract from) the agreement between the tracks and our results.  Notice that the stochastic errors considered in \S3.2 (again barring gravity; see further below) are also largely incapable of changing whether or not we lie on a specific mass-radius track, for the same reason: they directly affect only radius, thus moving objects parallel to the track locii.  

Second, the compatibility of our results with the theoretical mass-radius tracks depends primarily on how well our gravities agree with the predicted values.  Say our inferred log \gv for an object matches the (nearly constant) value predicted by the evolutionary models for its age.  Then, since our mass depends on our measured gravity through Newton's law, we are bound to fall {\it somewhere} along the mass-radius track for that age, regardless of the precise radius we derive.  Altering the radius will certainly change the inferred mass, but the object will still remain on the same track, which corresponds to a particular gravity.  Conversely, if our log \gv diverges from the model value for some age, we will also be offset from the corresponding mass-radius track at that age, no matter what radius we infer.

The above analysis reveals why all the targets whose log \gv were found to be consistent with the Lyon98/00 predictions for their assumed ages, in Paper I, also line up on the appropriate mass-radius locii in Fig. 1.  It is also clear why GG Tau Bb, USco 128 and USco 130 lie so far from the theoretical mass-radius locii: their inferred gravities are much lower than the model predictions for their expected ages (Paper I).  The crucial point is that uncertainties in radius (either systematic or stochastic) cannot alter these results\footnote{This analysis is certainly valid for the Upper Sco sample (estimated age 3--5 Myr): by $\gtrsim$ 3 Myr, the Lyon log \gv at a given age are indeed nearly constant with mass and \teff.  It is also valid for GG Tau Bb, even though the Lyon log \gv are {\it not} constant at $\sim$ 1 Myr: our gravity for Bb is less than the {\it lower limit} of Lyon values for this age (for any plausible mass or \teff; Paper I), so Bb must be offset from the Lyon 1 Myr mass-radius track regardless of the radius we derive.  GG Tau Ba is more complicated, and addressed in \S 4.2.  We can ignore it for now, since it does not appear discrepant on the mass-radius plot.}.  
   
Errors in gravity, on the other hand, which affect mass directly while influencing radius negligibly, can potentially be invoked to explain the radius discrepancy between our results and the Lyon tracks for the coolest, lowest mass targets.  Specifically, consistent underestimations in gravity for low \teff objects will shift them horizontally to spuriously low masses in Fig. 1, making their measured radii appear too large for their mass.  However, we have already argued exhaustively, both here (Appendix C) and in Paper I, against significant systematics in our log \gv; we do not consider this a likely explanation.  With respect to stochastic errors, it may be suggested that the radius discrepancies are evident only when we employ 1$\sigma$ (0.25 dex) uncertainties in log \gv; if, by fluke, our adopted gravities are in error by more than this, GG Tau Bb and USco 128 and 130 may be consistent with the tracks.  However, as noted above, what is really needed is a systematic underestimation of gravity.  It is very unlikely that this could arise by chance from random measurement errors, in all three of our coolest objects (all four, if one counts the less significant, but similarly directed deviation of USco 104 from the tracks).  Evidence that our stochastic errors in log \gv are small also comes from the absence of horizontal deviations from the evolutionary models at higher masses.  Since gravity hardly affects our radii, measurement uncertainties in log \gv will manifest themselves as large offsets in mass alone; no such appear at masses $\gtrsim$ 0.03 \msun (i.e., at higher masses, our gravities are remarkably consistent with the model predictions for the estimated ages; this is also apparent in the \teff/log \gv plot in Fig.9 of Paper I).  Deviations only at the lowest masses, and all in the same direction, argue strongly against large stochastic errors in log \gv.  

In conclusion, neither systematic nor stochastic errors in our analysis seem adequate to explain the disagreement in radius between the evolutionary model predictions and our values for the coolest, lowest-mass objects.  This prompts us to suggest that the evolutionary models may themselves be problematic.  We reached the same conclusion from our \teff/gravity analysis in Paper I (not surprising, since our compatibility with the mass-radius tracks depends on our agreement with the model gravities).  In both cases, using our parameters in conjunction with the theoretical tracks implies that the cooler (lower-mass) objects are much younger than the hotter (higher-mass) ones.  In Paper I, we argued that this age mismatch is spurious, both for our Upper Sco sample and (especially) for the two components of GG Tau B.  We further argued that the mismatch is caused by theoretical model uncertainties for the coolest objects, which is what we are also suggesting here as well.  The possible nature of these uncertainties was discussed in detail in Paper I; we touch on them again in \S 5.
  
\subsection{Radius and Mass versus \teff}
It might appear from the above discussion that at least the higher-mass objects ($>$30M$_J$), which agree with the Lyon98/00 models in both the temperature-gravity and mass-radius planes for their expected ages, are not a source of concern.  However, an inkling that all might not be well even in this mass regime comes from a consideration of GG Tau Ba.  

Compare the position of this object in the \teff-gravity plot (Fig. 9, Paper I) to its position in the mass-radius one (Fig. 1).  At the \teff we infer, GG Tau Ba lies quite far in gravity from the 1 Myr Lyon98/00 track in the \teff-log \gv plane; at the same time, it agrees very well with the 1 Myr mass-radius track.  This can only happen if our temperature for Ba is at odds with the evolutionary models: specifically, the gravity we derive is consistent with the Lyon98/00 1 Myr track only at a \teff $\sim$200K higher than our inferred value (Fig. 9, Paper I).  This effect is obvious for GG Tau Ba only because the theoretical tracks evince comparatively large variations in gravity with changing \teff at the youngest ages (due to initial condition effects, as discussed earlier).  In the Upper Sco targets, any such temperature offsets are largely masked in the \teff-gravity plane by the near constancy of log \gv at a specified age: objects can slide horizontally along the \teff axis while continuing to agree with the predicted gravity.  As we have shown in the last section, such agreement with the Lyon models in log \gv alone will also produce agreement with these models in the mass-radius plane, again without hinting at the underlying inconsistency in \teff.  In short, conflicts in temperature at Upper Sco ages cannot be probed via a simple comparison between the temperature-gravity and mass-radius planes, as is possible at the age of GG Tau.  We must explicitly compare mass and radius individually to \teff, as we now proceed to do.  

In Fig. 2 and Fig. 3, we plot our radii and masses against \teff, and compare to the Lyon98/00 tracks.  In both cases, substantial discrepancies are apparent between our results and the theoretical models for the higher mass objects ($>$ 30\mj).  For the temperatures we derive, their masses and radii appear significantly larger than the tracks suggest; equivalently, the models are hotter by $\gtrsim$ 200K than our values, for their inferred mass or radius.  Now, our four highest derived masses ($\sim$ 0.20--0.25 \msun; see Table 2) do seem slightly too high, given their spectral types (all our objects are M5 and later); in the field, 0.25 \msun corresponds roughly to M4, while M5 and later correspond approximately to masses $\lesssim$ 0.15 \msun (e.g., see Delfosse et al. 2000). Unrecognized binarity is an effect that will preferentially move objects to both larger radii and higher masses in our technique (Appendix E). If our highest mass objects were nearly equal-mass binaries, they would actually be $\sim$ 0.12 \msun, i.e., closer to the track predictions for their \teff. Since the translation between spectral type and \teff or mass has not really been tested for young low-mass objects, the viability of this possibility is a priori unclear; binarity should be checked through follow-up observations. 

It is hardly likely, however, that every one of our higher-mass objects is a binary. Even if this were true, the \teff discrepancy with the tracks would remain (though reduced to $\sim$ 100K).  Moreover, numerous studies show no evidence of binarity in GG Tau Ba (and our mass of 0.12 \msun is consistent with that of field objects of the same spectral type, $\sim$M6).  The fact that our \teff for this object also diverges from the Lyon models by $\sim$ 200K suggests that binarity is not the key problem here.  All this points to a real disagreement in temperatures.  

The Lyon98/00 evolutionary models use, as an outer boundary condition, significantly older versions of the Allard \& Hauschildt synthetic spectra, while we use an updated one.  It is known that the older generations of these spectra yield higher \teff for field M dwarfs than the newer ones do (e.g., compare the results of Leggett et al. 1996 to Leggett et al. 2000).  Moreover, the new spectra provide much better fits to the observed M dwarf spectral energy distributions (SEDs); the newer dwarf \teff thus appear more trustworthy \citep{Leggett00}.  Extending this to our PMS sample, the \teff we derive for our objects using the new synthetic spectra are also likely to be more accurate than any values derived using older versions.  This may explain the difference between our \teff and those in the Lyon98/00 models: the suggestion is that the Lyon models find hotter temperatures than our (presumably more accurate) values because they use older synthetic spectra.  

Although attractive, this suggestion is not necessarily correct.  The reason is that, while the evolutionary models indeed use the atmospheric calculations as an outer boundary condition, the actual \teff (and luminosity) implied by the models for a given mass is primarily a function of the interior calculations, and only slightly dependent on the atmospheric properties: the atmosphere in these objects is a very thin skin that is rather inconsequential for the evolutionary modeling.  For instance, much of the improvement in the newer synthetic spectra results from the use of updated opacities.  However, the temperature and luminosity evolution of these low-mass bodies is only marginally allied to photospheric opacity: $L$($t$) $\propto$ ${{\kappa}_R}^{{\sim}1/3}$, \teff($t$) $\propto$ ${{\kappa}_R}^{{\sim}1/10}$ \citep{Burrows93}.  Whether this slight dependence, combined with the difference in opacities between the older and newer spectra, is sufficient to give the $\sim$ 200K change in \teff we require, is doubtful.  Similarly, the newer synthetic spectra incorporate a more likely convective mixing-length parameter ($\alpha$) of 2 (as discussed in Paper I, $\alpha$$\approx$2 in both the upper and deep atmospheres is suggested by the latest 3-D hydrodynamic models for both PMS and field M dwarfs, while the older spectra and the Lyon98/00 tracks use $\alpha$=1).  However, BCAH02 show that this change in $\alpha$ appears important only for about the first million years, and hardly affects the Lyon models at later ages (for the masses of concern here), while our USco objects are supposed to be at $\sim$ 5$\pm$2 Myr.   

Thus, while the new synthetic spectra are crucial for our \teff determination from spectral analysis, and are hence instrumental in revealing uncertainties in the theoretical evolutionary tracks (assuming the spectral new spectral models are accurate), it is not clear that simply incorporating these new atmospheric calculations in the evolutionary modeling is by itself sufficient to remove the discrepancy between our results and the track predictions.  It is possible that the problem lies deeper, in the interior calculations of the evolutionary models.  

On the other hand, it is useful to examine the alternative hypothesis, that the Lyon98/00 temperatures are in fact correct, and it is our spectroscopically derived \teff that are amiss.  In Paper I, we have shown that the synthetic spectra employed reproduce the spectral features of our PMS sample remarkably well.  They also match the observed SEDs of field dwarfs much better than previous versions, as mentioned above.  However, these results do not, by themselves, rule out systematic offsets in the implied temperatures (e.g., due to inaccuracies in model opacities).  Is it possible that such systematics in our \teff are responsible for our conflict with the Lyon models?  We think not, for the following reason.  Our \teff are determined through a fine-analysis of the TiO bandheads, which are only marginally dependent on gravity.  Gravities, meanwhile, are deduced from the profiles of the absorption doublets of \na and \pot.  However, the latter lines depend sensitively on both temperature and gravity.  To fix log \gv uniquely, therefore, the temperature is fixed at the value implied by the TiO bandheads.  Consequently, if the model treatment of TiO is inadequate, leading to systematic offsets in the derived \teff, then we will derive the correct log \gv only if the synthetic spectra err in their treatment of the alkali lines as well -  and fortuitously by just the right amount, in both \na and \pot, to offset the independent error in \teff from TiO.  Given the completely independent parameters that enter into determining the behaviour of each of these species - TiO, \na and \pot - this would be quite a remarkable coincidence, and is not tenable.  It is far more likely that if our \teff are wrong, then our gravities are wrong as well.  However, we have seen that our gravities actually agree with the Lyon98/00 values for the higher-mass objects, for their estimated ages.  If our log \gv are incorrect in this mass and age regime, then it would appear so are the Lyon ones.  While such simultaneous offsets in both our and the Lyon values are possible, they are untestable without additional empirical constraints.  Moreover, this hypothesis simply replaces our original suggestion of a \teff error in the Lyon models with an error in Lyon gravities.  It is more reasonable to suppose that, when our independently derived log \gv match the Lyon predictions, both values are correct.  If so, it is hard to see how our spectroscopic \teff for the higher masses could be far off the mark.  It is worth noting in this regard that problems in the theoretical mass-\teff relationship are also emerging in stars that are somewhat higher in mass ($\sim$ 0.6--1 \msun) than those considered here, through studies of field and PMS eclipsing binaries \citep{Torres02, Stassun03}.  Various shortcomings in the evolutionary models, such as in the treatment of convection, have been proffered to resolve this issue (Stassun et al. 2003; Montalban et al. 2003).  It remains to be seen if similar effects are also applicable in the mass regime of interest here, and whether they can bring the Lyon temperatures into better agreement with our values.  
     
Finally, we point out that our temperatures for the lowest-mass objects (GG Tau Bb, USco 104, 128 and 130) are either consistent with, or {\it higher} than, the Lyon values in the mass-\teff plot, in contrast with the situation at the higher masses (Fig. 3).  In the radius-\teff plane, on the other hand, they appear cooler than the Lyon predictions for their radii, just like the more massive objects (Fig. 2).  This apparent contradiction in the behaviour of the low-mass bodies, when compared to the Lyon tracks - seemingly too cool in one parameter space, and too hot in another - clearly cannot be due solely to a temperature offset from the tracks.  Instead, its genesis lies in the fact, discussed earlier, that our gravities for these objects are incompatible with the Lyon values at any plausible \teff.  Specifically, their offset from the tracks in the radius-\teff plane can be attributed to their radii being too large for their inferred temperatures (just like their radii are larger than predicted for their mass).  Notice the subtle difference in interpretation of the radius-\teff plot for the higher and lower masses: in the former, the divergence from the Lyon tracks arises from a disagreement in \teff, while in the latter it arises primarily from a discordance in radius, even though the final result looks the same (all the objects appear younger than expected in the radius-\teff plane).  Meanwhile, the mass-\teff plot points to an {\it additional} offset in temperature at the two lowest masses (USco 128 and 130), which appear hotter than predicted for their inferred mass\footnote{Fig. 3 shows that uncertainties in our mass for these two objects can potentially reduce this \teff discrepancy, by bringing them closer to the position of GG Tau Bb.  However, we have already presented in \S 4.1.1 arguments in support of our derived mass, and are fairly confident that mass errors are not the problem here.  Notice, furthermore, that increasing their mass estimate, to agree more closely with that of GG Tau Bb, requires us to either increase the gravity at fixed radius, or increase the radius at fixed gravity.  This cannot alter, and can instead exacerbate, the offset of these objects from the Lyon radius-\teff tracks (just as GG Tau Bb is offset from the latter tracks, even though it agrees with the mass-\teff ones).}. 

\subsection{Mass versus Luminosity}
Finally, our mass-radius and mass-\teff results can be combined to examine the mass-luminosity relation in our sample (Fig. 4).  We saw that in masses $\gtrsim$ 0.03\msun, our radii agree with the Lyon98/00 predictions for the expected ages ($\sim$ 3-5 Myr for Upper Sco and $\sim$ 1 Myr for Taurus), while our \teff are lower than predicted for the same ages and masses.  At lower masses, our radii are much larger than predicted, while our \teff are roughly consistent with, or larger than, the predicted values.  Since $\lum$ $\propto$ ${\rad}^2$\teff$^4$, these results lead us to find luminosities that are slightly lower than indicated by the models for masses $\gtrsim$ 0.03\msun (for the expected ages), and substantially greater than predicted for lower masses.  

Notice that the luminosities for the higher masses are actually quite close to the theoretical ones, especially given our error bars (though they are slightly systematically lower than expected for the ages indicated by the mass-radius plot).  This is because our \teff disagreement with the models for these masses ($\gtrsim$ 200K), though large in absolute terms, constitutes a relatively small fractional discrepancy ($\sim$ 7\%).  This, combined with the good agreement with the models in radius, yields luminosities that are roughly in agreement with the theoretical ones.  At lower masses, however, our radii are a factor of $\sim$ 2 larger than in the models, resulting in a substantial divergence between our luminosities and the theoretical ones (this is compounded at the lowest masses - USco 128 and 130 - by our inferred \teff also being much larger than the predicted ones).  

Notice also that, in the Lyon evolutionary tracks, lower (higher) luminosity at a given mass corresponds to an older (younger) object.  This is because evolution in these models proceeds roughly along vertical Hayashi tracks over the first few Myr:  a given mass contracts with age (modulo deuterium-burning, which slows the contraction rate) at approximately constant \teff (see Fig. 2).  Our results show the same luminosity trend:  both components of GG Tau, which are expected to be younger than the Upper Sco objects, are more luminous than the latter.  This is a direct consequence of the fact that we find the GG Tau components to have a larger radius than similar mass but older Upper Sco targets (Fig. 1).  

\section{Conclusions}
Young star-forming regions contain a set of objects at similar distance, with similar ages and compositions. They are usually extensively studied photometrically, yielding colors. Combining photometry with spectra (which yield temperatures), one can determine extinctions and luminosities. These can then be leveraged to determine radii. Finally, high-resolution spectra can provide surface gravities, and thereby masses when combined with the radii. These stellar parameters, obtained with the aid of model atmospheres, can then be compared with those from theoretical isochrones, allowing a calibration of the fundamental evolutionary calculations that have been heavily used in the analysis of star-forming regions.

In a previous paper, we used ``fine analysis'' of high resolution-spectra of young, very low-mass objects to gain reasonably precise temperatures and gravities. In this paper, we use the photometry of these objects as described above to complete a measurement of their fundamental stellar parameters. Taking observational and model atmosphere errors and uncertainties into account, we reach two major conclusions:

{\it (1)} Both radius and \teff decrease less rapidly with diminishing mass, at a given young age, than predicted by the theoretical evolutionary models.  Specifically, in the mass-radius plane the lowest mass objects ($\lesssim$30 M$_J$) remain much larger (i.e., contract more slowly with age) than the models suggest, while the higher masses have radii in good agreement with the model predictions.  In the mass-\teff plane, the higher masses are substantially cooler than predicted, while the lowest masses have \teff either in better agreement with, or hotter than, the model values.  The combination of these two trends implies that luminosity also falls off less dramatically with mass, at a given age, than the evolutionary models indicate.  

{\it (2)} The lowest masses in our Upper Sco sample are near the deuterium fusion boundary.  

Because of the importance of both conclusions, we have taken considerable pains to consider possible sources of error, both observational and systematic. These include conversion of colors to extinctions, temperature scales for pre-main sequence objects, problems with the gravity measurements, and the effects of starspots or binarity. Our extinctions are consistent with an analysis of the same region using low dispersion spectra. Our temperature scale is in good agreement with recent photometric work in the field. The range of masses we find within a few spectral subclasses is perhaps surprising, but we show that some of our basic conclusions can be drawn just from the observations (without recourse to theory at all).  We conduct a comparative analysis with a more extensively-studied young (GG Tau B) binary system to further test our conclusions, and find comparable discrepancies with theory in that case as well.  Finally, our derived relationships between radius, mass, \teff and luminosity all agree (within the measurement uncertainties) with certain basic theoretical predictions that are likely to be correct regardless of evolutionary model uncertainties: younger objects have larger radii than older ones of the same mass; less massive objects are cooler and generally smaller than more massive ones at a given age; and luminosity decreases with both diminishing mass (at fixed age) and increasing age (at fixed mass).  There does not appear to be any a priori physical basis, therefore, for discarding our results.  We also point out that (like the Lyon models) our mass-radius, mass-\teff, radius-\teff and mass-luminosity relationships are smooth, without any sharp breaks or discontinuities.  The precipitous drop in gravity at low temperatures that we found in Paper I, which might seem remarkable at first sight, is due (if our analysis is correct) simply to a relatively slow change in radius and \teff (compared to the Lyon predictions) over a significant range in mass.  

The weight of the evidence suggests that substantially more work should go into the measurement of physical parameters of young substellar objects, the validity of the evolutionary tracks, and, without doubt, further testing and confirmation of our results.  An especially important conclusion of our work is that agreement with the evolutionary models in any single two-parameter plane (e.g., mass-radius) does not guarantee agreement in all parameters (e.g., \teff, luminosity).  In order to ascertain the veracity of the models, their predictions must be checked for all the parameters, not just a selected few.  As a corollary, comparing an object to the evolutionary models over one set of parameters (e.g., \teff-luminosity), in order to estimate other quantities (e.g., mass), is an exercise that is not always justified.  Such translations, which are common practice in current studies of young low-mass objects, may lead to spurious mass and radius estimates, and must be undertaken with great caution.  Similar conclusions have been reached by other authors, in the context of evolutionary model comparisons to higher-mass (solar-type) PMS stars (e.g., Torres \& Ribas 2002).
  
In Paper I, we pointed out some specific areas of concern for theory, such as accretion effects and the treatment of convection and deuterium fusion.  In particular, we noted that if deuterium fusion begin at an earlier time than predicted, the discrepancies in radius and gravity between the theoretical tracks and our measurements, for the lowest masses, may be resolved.  This is a testable hypothesis, as we outlined in Paper I, and bears closer examination.  In this paper, we have also identified discrepancies in the theoretical \teff predictions (assuming our derived temperatures are accurate), especially for the higher mass objects in our sample.  The underlying physical basis for temperature uncertainties in the evolutionary models is unclear; it is possible that remaining inadequacies in the treatment of convection are at fault.  Finally, while the model atmospheres and synthetic spectra that lie at the heart of our analysis are tremendously improved from earlier generations, they still suffer from certain shortcomings.  Specifically, they reproduce the photometry of field M dwarfs in some, but not all, of the optical and infrared bands.  While we have gone to great lengths to account for, and exclude, any attendant uncertainties in our analysis, further improvements in the atmospheric modeling - particularly in the linelists and opacities (most importantly, of H$_2$O) - would be tremendously useful for future studies of field and PMS low-mass objects.  

We have implemented methods that have long been used for normal stars. They provide a means of testing theoretical isochrones and obtaining fundamental stellar parameters for very young, very low-mass objects. This methodology (which highlights the importance of high resolution spectroscopy and model atmospheres) should also be extended to higher mass objects and other star-forming regions with different ages. Extensive programs of this nature are now both desirable and feasible.

\acknowledgments
We would like to acknowledge the great cultural significance of Mauna Kea for native Hawaiians, and express our gratitude for permission to observe from atop this mountain.  We would also like to express our thanks to the Keck Observatory staff, who have made possible, and successful, our observations over the last several years.  We would like to thank Russel White for kindly supplying two of the spectra used in this paper.  We would also like to thank Russel White, Lee Hartmann, Gilles Chabrier and Isabelle Baraffe for illuminating discussions on PMS evolution, and a constant readiness to help. S.M. would like to acknowledge the support of the SIM-YSO grant for his postdoctoral research. This work was supported in part by NSF grants AST-0205130 to R.J. and AST-0098468 to G.B.

\clearpage
\appendix
\section{Model Colors and Photometry}
To date, there are no empirical measurements of intrinsic colors and surface fluxes in cool, low-mass PMS objects, precluding any direct tests of the synthetic photometry in this regime.  However, Leggett et al. 2000 have recently derived temperatures for a number of similarly cool (but older) field M dwarfs with known distances.  We evaluate the reliability of the model photometry through comparisons with these objects, and assume, with appropriate caveats, that similar results hold in the cool M-type PMS regime.  

We have specifically analysed field dwarfs with solar metallicity, and \teff $\approx$ 2600--3000K (corresponding to M5--M6.5 types in Leggett et al's study), since abundances in our PMS objects are likely to be solar (Paper I), and their \teff from spectral analysis lie in the same range.  Moreover, we have focussed on behaviour in the $\rc$, $\ic$ and $J$ bands: the field and PMS objects are intrinsically too red for accurate shorter wavelength photometry, while the synthetic spectra are known to have some problems related to H$_2$O opacity in the $H$ and $K$ bands \citep{Leggett00}, the investigation of which is beyond the scope of this paper.       
    
Within these constraints, Fig. 5 shows that the synthetic spectra reproduce the field dwarf $\rc$-$\ic$ colors remarkably well: theory and observations agree to within $\pm$0.1--0.2 mag, with no evident systematic offsets.  $\rc-J$ and $\ic-J$ are worse, with the synthetic colors being clearly bluer by $\gtrsim$0.2 mag (extinction should be negligible for these nearby Main Sequence stars).  Thus, assuming a similar situation holds in the PMS regime, $\rc-\ic$ appears best suited to derive $\av$ for our targets; hence our use of this color index (eqn. [1]).  

On the other hand, Fig. 5 reveals that the {\it absolute} photometry of the field dwarfs is best reproduced by the synthetic spectra in the $J$ band.  The models are too bright in $\rc$ and $\ic$ by upto $\sim$ 0.5 mag, but only by $<$0.2 mag in $J$.  Indeed, this appears to explain the color behavior noted above: the overluminosity of the models in $\rc$ and $\ic$ compared to the observations, combined with the more accurate $J$-band predictions, makes the synthetic $\rc-J$ and $\ic-J$ colors too blue; simultaneously, the errors in model $\rc$ and $\ic$ fluxes, similar in magnitude and direction, cancel to make the $\rc-\ic$ predictions commensurate with the data.  At any rate, the synthetic $J$-band fluxes seem most appropriate for estimating the true surface flux, and hence radius, in M-type field dwarfs, and by extension in our M-type PMS sample.  Using model $\ic$ (or $\rc$) to calibrate the true surface flux (once extinction is corrected for using $\rc-\ic$) will, if the field dwarf results apply to the PMS regime, cause us to underestimate radius, and hence mass and luminosity (basically, we will assume that the stellar surface is brighter per unit area than it really is, in $\rc$ or $\ic$, thus requiring a spuriously low surface area, or radius, to produce a given $\av$-corrected observed flux).  In fact, the synthetic $\ic$ fluxes do appear overluminous compared to $J$ in the PMS regime: our $\ic$-radii are systematically somewhat lower (on average by $\sim$20\%) than the $J$ ones (\S 4).  

However, there is also some evidence that the synthetic $\rc-\ic$ colors, while accurate for the field dwarfs, suffer from systematic offsets for our PMS sample.  On the one hand, our inferred Upper Sco extinctions are commensurate with those found by Preibisch et al. 2002 for their Upper Sco PMS targets of similar spectral type ($\av$ derived using a method complementary to ours). However, the Upper Sco region is known to have very little remaining nebulosity, and 100$\mu$m surveys towards the region indicate an average $\av$ $\sim$ 0.5 mag, about 1 mag less than our mean value.  Nebulosity on small spatial scales, surrounding the individual stars, might account for this discrepancy.  Alternatively, it is possible that the PMS model $\rc-\ic$ are systematically too blue by $\sim$0.2 mag, leading us to overestimate $\av$ by $\sim$1 mag (i.e., $\ai$ and $\exj$ by $\sim$ 0.6 and 0.3 mag respectively).  

Such an offset in extinction would have the following consequence.  If our synthetic $J$-band surface flux estimates are accurate (as the field dwarf results suggest), then overestimating $\exj$ by 0.3 mag would produce corresponding overestimations in radius, mass and luminosity (15\%, 30\% and 30\% respectively).  Conversely, if our model $\ic$ surface fluxes are too high (again as indicated by the field dwarfs), combining them with the erroneous extinctions would actually yield reasonably {\it correct} radii, masses and luminosities: the proposed systematic error in $\ai$ ($\sim$0.6 mag) is about equal to that in $\ic$ flux ($\sim$0.5 mag in the field), but the two offsets act in opposite directions  - higher $\ai$ implies larger radius, while higher $\ic$ flux at the stellar surface implies lower radius - and thus largely cancel out (eqn. [3]).    

To summarize: field dwarf comparisons indicate that the synthetic $\rc-\ic$ colors and $J$-band surface fluxes are accurate, while the model $\ic$ surface fluxes are too high.  Assuming this holds for our PMS sample, our extinctions and $J$-based radii, masses and luminosities are accurate, while the $\ic$-based values are underestimations.  However, our extinctions may be too high due to model $\rc-\ic$ offsets in the PMS regime.  In this case, the values inferred using $\ic$ fluxes continue to be lower than those from $J$, but the $\ic$-based parameters are in fact more accurate, while the $J$ ones are overestimations.  We cannot currently distinguish between the two possibilities.  However, the $\ic$ and $J$ calculations should reasonably bracket the true values.  We thus provide both sets of estimates (Table 2); our primary conclusions (\S4) remain unaltered independent of which set is adopted.  In fact, in the majority of cases our parameters from $\ic$ and $J$ are very similar (Table 2): radii agree to within $\sim$ 15\%, and thus masses and luminosities to within $\sim$ 30\%.  In the few cases where this is not true, errors in the observed optical photometry are likely to blame, as addressed in Appendix B.  

Finally, the possible overluminosity in the synthetic $\ic$ fluxes may raise some questions about the validity of our \teff and log \gv inferred from spectral analysis (Paper I), since most of our spectral diagnostics lie in the $\ic$ band.  This issue is addressed in Appendix C.  

It is worth pointing out here the pitfalls associated with analysing PMS low-mass objects based on spectral type considerations alone, as is common practice in current research.  Any such analysis requires assumptions about intrinsic PMS colors and photometry, in order to derive extinctions, luminosities and so on.  As Fig. 5 shows, colors are strongly dependent on \teff and (to a lesser extent) gravity.  Thus, using spectral types to derive PMS parameters requires an accurate translation from spectral type to \teff and log \gv.  Such a translation, however, is at present sorely underdeveloped.  Most investigators assume field M dwarf colors for low-mass PMS objects; however, the spectral-type to \teff conversion for field dwarfs remains uncertain by roughly $\pm$ 100K (e.g., Leggett et al. 2000). Moreover, PMS gravities are considerably lower (by 1--2 orders of magnitude) than those of field objects.  Both effects can lead to significant errors in assigning PMS colors (see Fig. 5).  More importantly, it is not at all clear that PMS temperatures are the same as that of field dwarfs, for a given spectral type.  Indeed, the most widely adopted PMS \teff scale these days is that of Luhman 1999, who advocates PMS \teff systematically higher, by $\sim$ 100--200K, than in field dwarfs of the same spectral type.  This PMS \teff scale is based on the {\it requirement} that PMS observations agree with the Lyon98/00 theoretical evolutionary tracks, and is thus completely model-dependent.  Nevertheless, assuming it is qualitatively correct (i.e., that PMS \teff are higher than dwarf ones), it is obvious that field dwarf colors are not appropriate for the PMS domain.  In particular, many analyses of M-type PMS objects simultaneously assume {\it (1)} intrinsic colors similar to that of field dwarfs (for calculating extinctions and luminosities), and {\it (2)} the Luhman 1999 \teff scale (for putting the PMS objects on an H-R diagram).  The concurrent adoption of both assumptions is internally inconsistent, and untenable.  

The attraction of using spectral types for PMS analysis, clearly, lies in the ease with which types can be determined.  Nevertheless, spectral types are a purely empirical construct.  In order to employ them profitably, it is imperative to derive a priori the connection between types and physical conditions.  In particular, the above discussion shows that one must first establish (without recourse to evolutionary model predictions) a spectral type to \teff conversion scale for the PMS regime.  Detailed spectral analyses, such as undertaken in Paper I, are necessary to accomplish this.  Of course, our present work (Paper I and here) includes only a small sample, and covers a very limited range in spectral types.  It is thus insufficient to derive a robust spectral type - \teff scale for PMS objects (nor is the derivation of such a scale our intent in this work).  Future studies with larger samples must address this issue.  The crucial point, however, is that our analysis eschews spectral type considerations; we explicitly derive \teff and log \gv for our PMS targets, and then use the appropriate colors and photometry (based on model atmosphere calculations).  In so doing, we avoid the current uncertainties associated with spectral type to temperature, gravity and color conversions.  

\section{Errors in Observed Photometry}
In three of our targets (USco 100, 109, 128), the $J$-band radii are more than 20\% higher than the $\ic$-band ones; in two others (USco 112 and 130), they are more than 30\% higher.  This is surprising, given that in all the remaining objects the discrepancy is $\lesssim$ 15\%, and mostly less than 10\% (Table 2).  It cannot be due to synthetic photometry problems that increase with changing \teff or gravity: the temperatures and gravities of the five anomalous objects span the range derived for our entire sample, and other targets at these \teff and log \gv present no difficulties.  A closer look reveals that these five are also our faintest targets (Table 1).  This leads us to propose that the disagreement between their $\ic$ and $J$ values is due to systematically larger errors in their observed optical photometry (i.e., in $\rc$ and/or $\ic$, adopted from AMB00) than in the other, brighter objects.  This suggestion is driven by the appearance of color anomalies in these faintest targets, as we now show.

For the sake of concreteness, we compare USco 130 (which exhibits the largest radius anomaly: a 40\% difference between $\ic$ and $J$) and GG Tau Bb (whose $\ic$ and $J$ radii are nearly identical).  From spectral analysis (independent of reddening), we have found nearly identical \teff for the two; regardless of any systematics in our precise \teff value, the equivalence of their temperatures is robust, given the close similarity between their TiO bands (which are highly sensitive to \teff differences and negligibly to gravity; Paper I).  Thus, since photometric colors depend predominantly on \teff and very marginally on gravity (which we find to be nearly the same in both anyway), we expect their intrinsic colors to be very similar.  However, this expectation is not borne out.  After accounting for extinction, USco 130 is fainter than GG Tau Bb by $\sim$ 2 mag in $\rc$ and $\ic$, but by only $\sim$ 1 mag in $J$.  In other words, its $\av$-corrected $\rc-\ic$ color is the same as Bb's (which is guaranteed since we derive $\av$ from $\rc-\ic$), but its $\ic-J$ and $\rc-J$ are much redder.  Analogous discrepancies occur regardless of which pair of bands $\av$ is calculated from.  The problem can be traced directly to the observed photometry (Table 1).  We see that USco 130 and Bb have exactly the same observed $\rc-\ic$ (2.46 mag), but USco 130 appears far redder in $\ic-J$ (3.16 vs. 2.39 mag) and $\rc-J$ (5.62 vs. 4.85 mag).  

This behavior is very difficult to recreate through physical effects.  Given two stars with the same \teff, extinction differences cannot redden $\ic-J$ and $\rc-J$ and leave $\rc-\ic$ unchanged.  Ultra-cool spots, by contributing increasing flux with longer wavelength, can potentially be responsible.  In reality, however, the $\rc$, $\ic$ and $J$ bands are too close together for this to produce any large effect: spots too cool to contribute much flux in $\rc$ and $\ic$ compared to the photosphere (thus leaving $\rc-\ic$ unchanged) also do not yield significant flux relative to the photosphere in $J$ (and so do not change $\ic-J$ or $\rc-J$ much either).  The same is true for cooler companions, whose effect is akin to that of spots (Appendix D).  Another alternative is the presence of excess NIR flux from a disk.  However, such emission should be minimal in $J$.  Moreover, there is no evidence for any substantial circumstellar material around USco 130: neither any high-resolution spectral signatures of disk-accretion (JMB02), nor any excess emission even at longer NIR wavelengths ($K$-$L'$), where it should be far more evident than at $J$ \citep{Jaya03}.  Finally, our $J$-band photometry for the USco sample is from 2MASS; the errors in this are $\lesssim$ 0.03 mag, much too small to produce the above effects.  It is thus safe to conclude that a spuriously high $J$ flux, engendered either by real phenomena or errors in observed $J$, cannot account for USco 130 being redder in $\ic-J$ and $\rc-J$ than GG Tau Bb.  At the same time, uncertainties arising from synthetic photometry errors cannot be responsible.  Since the observed $\rc-\ic$ is the same in both, and so is their \teff, their implied $\av$ must also be very similar, independent of the precise extinction we derive using model colors.  If the $\av$ is the same in both, then the other observed colors ($\ic-J$, $\rc-J$) should also be the same, for photospheres at the same \teff.  The only remaining solution is that the observed $\rc$ and/or $\ic$ photometry for USco 130 is incorrect.  This can lead to the sort of effects we see, as illustrated shortly.

A comparison of the $H$ and $K$ photometry for USco 130 (from 2MASS) and GG Tau Bb (from WGRS99) supports the above hypothesis.  The observed values are $H$ = 13.54, $K_s$ = 13.08 for USco 130, and $H$ = 12.38, $K$ = 12.01 for Bb (the intrinsic offset between the $K_s$ and $K$ filters is negligible for our purposes and can be ignored; Carpenter 2001).  We see that in both filters, USco 130 is fainter than GG Tau Bb by $\sim$1.1 mag, a similar reduction in flux as in the observed $J$ band (Table 1).  Correcting $H$ and $K$ by our derived extinctions does not change this result: USco 130 remains $\sim$ 1 mag fainter than Bb in both bands, just like in $\av$-corrected $J$.  The fact that accounting for extinction does not affect the flux difference between USco 130 and Bb is of course not surprising, given that our $\av$ for both is nearly the same.  However, the point is that the difference between the two objects in $H$ and $K$ agrees with that in $J$, and not with the $\sim$ 2 mag difference in $\rc$ and $\ic$.  Since all the arguments above supporting the accuracy of the $J$ photometry are equally applicable to $H$ and $K$, we are once again led to conclude that the observed $\rc$ and $\ic$ values are incorrect.  

Note that, since USco 130 is fainter than Bb by similar amounts in all three NIR bands, their observed NIR colors are very alike.  As we find their \teff to be the same, this might suggest that their extinctions are also identical, just as implied by the equivalence of their $\rc-\ic$ colors.  In that case, the $\rc$ and $\ic$ fluxes quoted by AMB00 would both have to be underestimations by nearly 1 mag, in order to obtain the same flux difference in the optical as in the NIR without altering the derived $\av$.  This is implausible.  However, $J$, $H$ and $K$ are actually very insensitive to extinction, compared to $\rc$ and $\ic$.  Thus for any reasonable variation in $\av$ between USco 130 and GG Tau Bb (see below), the observed differences in their NIR colors will still closely reflect their intrinsic differences.  Thus, what their resemblance in NIR colors really implies is that their \teff are indeed very similar, as we claim.  Small variations in $\av$ and reasonable errors in $\rc$ and $\ic$ are still perfectly admissible, as we now show.  

With the available information, it is impossible to uniquely determine the errors in observed $\rc$ and $\ic$.  However, we can still run a plausibility check by considering likely values.  Imagine that the $\rc$ magnitude quoted by AMB00 for USco 130 is too high (i.e., $\rc$ flux too low) by 0.3 mag, and the $\ic$ too high by 0.4 mag.  If so, USco 130 is also 0.1 mag redder in $\rc-\ic$ than the AMB00 value; our extinction estimate must therefore increase by 0.47 mag (bringing the total $\av$ difference between USco 130 and Bb to 0.47+0.14 = 0.61 mag; see Table 1).  Correcting for these offsets produces a difference of 1.1 mag between USco 130 and GG Tau Bb, in both $\rc$ and $\ic$.  At the same time, the change in $\av$ alters our previous estimates of intrinsic $J$, $H$ and $K$ by $\lesssim$ 0.1 mag; USco 130 then remains $\sim$1 mag fainter than Bb in the NIR.  Thus, the optical and NIR differences between the two are now completely consistent.  Notice that AMB00's quoted errors in $\rc$ and $\ic$ are $\lesssim$ 0.1 mag.  However, they cite no increase in errors with decreasing brightness, which seems rather unrealistic given that their sample covers $\sim$ 5 magnitudes in $\rc$ and $\ic$, with the faintest objects (including the anomalous ones discussed here) lying near or below their completeness limit in $\rc$ and $\ic$ ($\sim$ 19 and 18.5 respectively).  We think it quite within the bounds of reason, therefore, to postulate 0.3--0.4 mag errors in $\rc$ and $\ic$ for USco 130, which is the faintest target in our sample (and among the very faintest in AMB00's).  Of course, the above exercise is not proof that these are indeed the precise errors in $\rc$ and $\ic$.  It serves to demonstrate, however, that {\it (1)} plausible uncertainties in the observed optical photometry can easily explain the discrepancy between the optical and NIR bands in USco 130, and {\it (2)} the $J$ photometry is likely to be more accurate (i.e., not subject to significant change upon correcting for these uncertainties), and thus the $J$-based radius more trustworthy (notice that, once the optical photometry is corrected in the illustrative exercise above, the $\ic$ radius, as expected, becomes consistent with the $J$ one, just as it is in GG Tau Bb).   

Similar arguments can be made for USco 128.  From 2MASS, we have $H$ = 13.78 and $K_s$ = 13.21 for this object.  Its observed difference with Bb ($H$=12.38, $K$=12.12.01; WGRS99) is then 1.40 and 1.20 mag in $H$ and $K$ respectively, very similar to the 1.25 mag difference in observed $J$ (Table 1).  Accounting for our derived extinctions yields intrinsic $H$ and $K$ differences of 1.53 and 1.28 mag; once again, consistent with the 1.5 mag intrinsic difference we derive in $J$.  For USco 130, we have illustrated that correcting for discrepancies between its (erroneous) optical and (more accurate) NIR photometry probably leads to a change of $\lesssim$ 0.1 mag in the intrinsic $J$, $H$ and $K$ derived using $\av$ based on AMB00 photometry.  In USco 128, the divergence between the optical and NIR photometry is even smaller than in 130\footnote{The apparent difference between USco 130 and GG Tau Bb in the optical, after $\av$ correction, is 2 mag, while in the NIR it is 1 mag; this is the discrepancy we ascribe to errors in AMB00's photometry for USco 130 (\S 3.3.2).  In USco 128, the corresponding values are 2 mag and 1.5 mag, so the optical and NIR values are divergent by only 0.5 mag, implying a smaller correction to the AMB00 photometry for the latter object.}, so the intrinsic $J$, $H$ and $K$ we derive for it, and the corresponding offsets from GG Tau Bb in these filters, should be quite accurate.  In other words, $H$ and $K$ photometry, which is even less affected by $\av$ errors than $J$, supports the claim that USco 128 is 1.5 mag fainter than GG Tau Bb; if we conservatively adopt the observed photometry without any $\av$ correction, the difference between the two in the NIR is still $\sim$ 1.3 mag (implying 11 \mj for USco 128, close to our adopted estimate of 9 \mj and still in the planemo regime).      

Analogous conclusions can also be drawn for the three remaining anomalous objects, by comparing them to targets at similar \teff and gravity that do not exhibit large differences between $\ic$ and $J$ radii (e.g., compare the discrepant object USco 112 to USco 75).  Specifically, their observed $\rc$ and $\ic$ fluxes from AMB00 appear too low.  It is this underluminosity (combined with attendant offsets in $\av$), which we claim is spurious just as in USco 130, that ultimately results in their $\ic$ radii (and hence masses and luminosities) being significantly lower than the $J$ ones.  Note that, since the divergence between $\ic$ and $J$ radii is smaller in these targets than in USco 130, the required corrections to their observed $\rc$ and $\ic$ are also correspondingly lower.  At any rate, the implication, as in USco 130, is that the parameters derived from their $J$ photometry are more dependable than the $\ic$-based ones.  In all the other targets, using $\ic$ or $J$ makes little difference.      

\section{\teff and Gravity}
It may be that our \teff and log \gv, inferred from detailed comparisons to synthetic spectra, are systematically erroneous (due to systematics in the model spectra).  In Paper I we argued that these errors should be quite small.  Nevertheless, it is fruitful to examine the consequences of such offsets for our present calculations.  In the last section, we found evidence for some systematics in the synthetic photometry, at least in the field M dwarf regime.  In the present analysis we assume that, notwithstanding any such absolute offsets, at least the differential model photometry is correct: i.e., that the models accurately predict the {\it change} in colors and fluxes for a given shift in \teff or gravity.  Under this assumption, we first investigate the effect of systematic \teff and log \gv offsets on our derived extinctions, radii and so on.  We then discuss the feasibility of such systematic errors in temperature and gravity, in light of the synthetic photometry results discussed in the last section.  

In \S 4.2, we show that our \teff are generally lower than the evolutionary track predictions, by up to 100--200K.  {\it If} this is due to an underestimation in our \teff (and we argue in \S4.2 that it is not), then our adopted synthetic photometry must be altered as well, to reflect the new temperatures; we wish to calculate the resulting change in our other parameters.  Now, every 100K rise in \teff (in the \teff range of interest here, 2500--3000K) produces a decrease in model $\rc-\ic$ by $\sim$ 0.1 mag, and a decrease in model $\ic$ and $J$ magnitudes by $\sim$ 0.2 and 0.15 mag respectively; i.e., the star is predicted to become both bluer in $\rc-\ic$, and brighter in $\ic$ and $J$.  For a given observed $\rc-\ic$, the change in model color implies $\av$ larger by $\sim$ 0.5 mag, i.e., an increase in $\ai$ and $\exj$ by 0.3 and 0.15 mag respectively.  However, as discussed in \S 3.3.1, simultaneous increases in both extinction and estimated surface flux act in opposite senses in the radius calculation (eqns. [3a,b]): the net change in inferred radius is then very small.  For a 100--200K rise in \teff, the numbers above imply an increase in radius by only 5--10\% when $\ic$ fluxes are employed, and a corresponding increase in mass and $\lum$ by just 10--20\%; with $J$ fluxes, the changes are even less.  These offsets are clearly insignificant compared to our adopted stochastic errors of 30\%, factor of 2, and 65\% in radius, mass and luminosity.  We are thus confident that our values for the latter parameters are largely unaffected by the photometric effects of any plausible systematic offset in \teff.  Notice, however, that since we calculate luminosity through $\lum$ $\propto$ $\rad$$^2$\teff$^4$, a 100--200K rise in \teff will directly raise our $\lum$ estimates by a further $\sim$ 15--30\% (in addition to the modification discussed above, which arises due to the photometry-related change in radius).  

Similarly, we have found large (0.50--0.75 dex) variations in gravity within our sample, at odds with the evolutionary track predictions (Paper I).  If our log \gv are are in error, then the attendant offsets in adopted synthetic photometry might lead to errors in our other parameters as well.  However, it turns out that changes in gravity produce very small offsets in the synthetic photometry.  For instance, even a 1 dex offset in our log \gv alters the model $\rc-\ic$, $\ic$ and $J$ by $<$0.1, $\lesssim$0.1 and $<$0.05 mag respectively.  Our radii, and hence masses and luminosities, are thus negligibly affected by gravity-induced photometric errors.  

However, gravity offsets will certainly affect mass directly and strongly, through our use of Newton's law to derive mass, independent of gravity-related photometric effects (which affect mass via tiny changes in radius).  In Paper I, we have already argued extensively against large systematic uncertainties (greater than our measurement errors of $\pm$ 0.25 dex) in our inferred log \gv, based on both a detailed evaluation of the synthetic spectra as well as purely empirical inter-comparisons of the observed spectra.  In particular, our tests there strongly suggest that the large gravity variations in our sample are due neither to physical effects such as dust, cool spots and metallicity fluctuations, nor to problems in the synthetic spectra arising, for example, from an inadequate treatment of collisional broadening.  However, we showed in \S 3.3.1 that the synthetic $\ic$ fluxes appear systematically higher than observed in field M dwarfs; by extension, such an effect may also occur in our PMS sample.  Since most of our spectral diagnostics in Paper I lie within the $\ic$-band, it is instructive to dwell for a moment on the implications of any such overluminosity for our \teff and gravity results.  

For a given \teff, the surface flux per unit area, $\sigma$\teff$^4$, is fixed; too much flux in any spectral region must then be compensated for by too little in others. An overluminosity in $\ic$ then implies that either {\it (1)} the model continuum opacity in $\ic$ is too low, allowing too much flux to escape in this bandpass (and correspondingly too little in others), or {\it (2)} the opacity in some other bandpass(es) is artificially high (leading to the same effect).  It is the first possibility that is a cause for concern, since synthetic opacity problems in $\ic$ can cause errors in our \teff and log \gv, derived from modeling of spectral lines within this bandpass.  Specifically, TiO bandheads form our main temperature diagnostic; this molecule is also the main continuum opacity source in the optical for M-type objects.  If the synthetic spectra underestimate the TiO opacities, then our \teff estimates will be too low (because the model bandheads at a given \teff are too weak, forcing us to choose cooler temperatures to match the data).  Concurrently, our gravities will also be spuriously low: at the low \teff inferred from the TiO bandheads, the \na doublet (our main gravity diagnostic) will be too strong (since \na strength depends on both \teff and gravity; Paper I), and we will adopt an artificially low gravity in order to match the observed doublet profile (assuming the model treatment of \na itself, which contributes negligibly to the continuum opacity, is correct).    

The above is a qualitative argument.  It is by no means certain, though, that erroneous TiO opacities will actually permit simultaneous good fits to both the \na doublet as well as the surrounding continuum, as we obtain in Paper I.  Secondly, fits to \pot give us gravities consistent with those from \na; since \pot and \na differ in their \teff/gravity dependencies, this is unlikely to happen if our temperatures are significantly inaccurate.  Thirdly, the M dwarf comparisons show marked $\ic$ overluminosity over our entire PMS \teff range; as such, we should underestimate gravity for all our targets.  However, our log \gv disagree substantially with the evolutionary tracks only for the cooler objects; the hotter ones agree very well with the tracks.  Fourthly, as a corollary, we should not find gravity variations between objects at the same temperature (i.e., with similar TiO bandhead strengths), which is nevertheless seen in our sample (Paper I).  These considerations strongly indicate that the overluminosity in synthetic $\ic$ is not due to opacity problems in the $\ic$-band itself, but in other bandpasses (see below).  In this case, our \teff and log \gv should be accurate: even if excessive flux is pushed out over the entire $\ic$-band, our spectral analysis, dependent only on the relative interplay between continuum and various line opacities within $\ic$, remains unaffected.   

In this context, it is noteworthy that significant discrepancies remain in the model treatment of H$_2$O, which is the main source of continuum opacity in the near-infrared.  The linelists appear incomplete, and disagreements are apparent between the models and observed low-resolution NIR spectra \citep{Allard00, Leggett00}.  Thus, given the above arguments, and the remarkably good and consistent fits we obtain to our $\ic$-band spectral diagnostics, we think it far more probable that any overluminosity in the $\ic$ bandpass is due to an inadequate treatment of H$_2$O in the NIR (namely, an overestimation of H$_2$O opacity, probably in the $H$ and/or $K$ bands), and not problems in $\ic$ opacities.  Consequently, we expect our \teff and log \gv values to be reasonably accurate, without substantial systematic offsets.
    
\section{Cool Spots} 
We now consider the possible effects of cool surface spots on our analysis.  The main result is that cool spots, covering up to 50\% of the stellar surface and cooler than the surrounding photosphere by up to 500K, do not affect our results significantly.  Such spots causes us to underestimate \teff from spectral synthesis by at most $\sim$ 200K (see also Paper I for a more detailed discussion of this); for our targets, with \teff in the range $\sim$ 3000--2500K, this leads to an `underestimation' of luminosity by $\sim$25\% (whether this is a true error is a matter of taste, since such a large cool spot does lower the luminosity below the value expected from an unspotted, hotter photosphere).  Moreover, the extinction, radius and mass estimates are affected negligibly. Fundamentally, these quantities depend on the {\it difference} in \teff derived from the spectral synthesis and from the photometry, while the presence of a spot (cooler than the photosphere by $\lesssim$ 500K) affects both the spectra and the photometry similarly.  Finally, spots that are either closer in temperature to the photosphere, or smaller than we assume, affect our results even less.

However, a large {\it ultra}-cool spot could affect our conclusions.  Such a spot would be completely dark against the stellar surface and contribute insignificant flux; thus, its effect on \teff from spectral synthesis, and on extinction, would be negligible.  However, if its areal coverage is a good fraction of the stellar surface, the star will be fainter than in the absence of the spot, leading us to underestimate radius and mass.  For a dark spot covering 50\% of the surface, our inferred radius would be too small by a factor of $\sqrt{2}$, and mass too low by a factor of 2; for a more plausible 30\% coverage, the respective underestimations are $\sim$15\% and 30\%.  In \S 4.1.1, we address this scenario with respect to some of our targets.  

We reach the bove conclusions by modeling the effect of spots using the synthetic spectra predictions for various photospheric temperatures.  We consider only differences in \teff between the cool spot and the surrounding photosphere, and neglect any differences in gravity (see Paper I for a discussion of those).  This approach is vindicated by the synthetic spectra, which indicate that changes in \teff affect the photometric fluxes and colors much more than changes in gravity do; it is also supported by comparisons between giant and dwarf photometry by other authors (e.g., Gullbring et al. 1998).  We will concern ourselves here only with $\rc$, $\ic$ and $J$ band photometry, since these are the bands we use for our extinction, mass and radius analyses.  We will also assume that both the observed photometry and the high-resolution spectra are affected by spots with similar \teff and covering fraction, despite the fact the photometry for our targets was obtained a few years before the spectra (this is further discussed at the end of this Appendix).  For the sake of argument, we assume the unspotted photosphere to have \teff = 3000K and log \gv = 4.0 (as in the spot discussion in Paper I); the results for other \teff and gravities within the range of our interest (\teff $\sim$ 2500-3000K, log \gv $\sim$ 3.0-4.0) are very similar.  We first consider spots cooler by 500K than the surrounding photosphere (i.e., spot \teff = 2500K), and covering half the stellar surface; we then discuss the effect of much cooler spots. We also recall here, from the spot discussion in Paper I, that under these assumed conditions (3000K photosphere, 2500K spot, log \gv=4.0, covering fraction=50\%), our {\it spectral} analysis of Paper I would imply \teff $\approx$ 2800K.  This fact will become useful in the following discussion.
   
For the \teff assumed for the spot and unspotted photosphere, the peaks of their spectral energy distributions are in the NIR, and the $\rc$ and $\ic$ bands lie in the Wien part of their spectra.  Consequently, the cooler spot contributes more flux in $\ic$ than in $\rc$.  This immediately implies that the intrinsic $\rc$-$\ic$ color of the spotted star will be that of an object cooler than than the photospheric temperature of 3000K (but no cooler than the spot temperature of 2500K).  For a rough a priori estimate (checked below through detailed examination of the synthetic spectra), we may assume that the $\rc$, $\ic$ and $J$ fluxes scale as the bolometric flux.  In this case, with the spot covering half the surface, the resulting flux in all three bands (and thus the $\rc$-$\ic$ color as well) will be similar to that from an object with \teff $\approx$ 2800K (= [[3000$^4$ + 2500$^4$]/2]$^{1/4}$).    

These expectations are confirmed by our analysis of the photometry predicted by the Allard and Hauschildt models.  We combine the model fluxes for a 3000K and a 2500K object, both at log \gv = 4.0.  We find that the resulting $\rc$, $\ic$ and $J$ band fluxes correspond to \teff of $\sim$ 2825, 2800 and 2775K respectively.  The intrinsic $\rc$-$\ic$ color of the spotted star is found to correspond to \teff $\sim$ 2875K.  Note that these photometric temperature estimates are very similar to the \teff$\approx$2800K that would be derived for the same spotted star from our spectral analysis of Paper I.  What does this imply for our extinction, mass and radius analysis?  As detailed in \S 3, we infer $\av$ by comparing the $\rc-\ic$ color implied by our derived \teff and gravity (from spectral fits), to the observed color.  We then derive radius by combining the observed $J$ magnitude, synthetic $J$ magnitude expected at the stellar surface (given our derived \teff and gravity), $\av$ and distance.  Consequently, our derived $\av$, radius and mass will be affected by cool spots, only if there is a substantial {\it difference} between the \teff inferred from our spectral fits and that suggested by the intrinsic photometry ($\rc-\ic$ color and $J$ magnitude).  

The results above show that the \teff difference is negligible, even for a large spot that is cooler than the surrounding photosphere by $\sim$ 500K.  The spectral lines and $\rc-\ic$ color are almost identically affected by the spot; the \teff inferred from them differ by only $\sim$75K (which is of order the $\pm$50K uncertainty in our \teff determinations anyway).  According to the models, this leads to a difference between intrinsic and assumed $\rc-\ic$ color of only 0.08 mag, and hence an error in $\av$ of only $\sim$ 0.4 mag.  Even without accounting for cool spots, our uncertainties in \teff and gravity determination and in the observed photometry lead to $\av$ errors of $\sim$ 0.7 mag (\S 3.3).  Moreover, most of our derived $\av$ are $\gtrsim$ 1 mag, much larger than the cool spot effect.  For both reasons, cool spots affect our $\av$ results negligibly.  

Now the radius determination involves, apart from $\av$ (and distance), the observed $J$-band flux and that expected from our derived \teff.  However, we showed above that (modulo distance) the intrinsic $J$-band flux will be the same as that expected from our spectral fits (i.e., both indicate almost exactly the same \teff) even when cool spots are present, so differences in $J$-band flux due to cool spots can be neglected as an additional source of error for radius determination.  The $\av$ errors due to cool spots {\it will} contribute, but marginally so as discussed above, since they are much less than both our $\av$ uncertainty due to other causes, as well as the absolute $\av$ values we find without considering cool spots.  The same holds true for our mass derivation.  Note here that we have also used $\ic$ fluxes, instead of $J$, to compute an alternate set of radius and mass estimates for our sample (see \S 3.3 and Appendices A and B).  However, the same arguments apply to $\ic$ fluxes as well: the intrinsic $\ic$ flux of the spotted star corresponds to the same \teff as derived from our spectral fits, so the presence of the spot does not affect our $\ic$-based radii and masses.  To summarize, we do not expect even large spots, cooler than the surrounding photosphere by $\sim$ 500K, to affect our $\av$, radius and mass derivations to any significant degree.  Hotter or smaller spots, of course, will affect us even less, since they will imply a \teff, in both observed photometry and in the spectral lines, that is even closer to that of the unspotted photosphere.

What about much cooler spots?  Since they will contribute negligible flux compared to the hot photosphere, they will not affect our \teff, gravity and $\av$ determinations.  However, if they cover a large fraction of the stellar surface, the observed $\ic$-band flux will be much less than that in the absence of the spot.  If the spot covers half the surface, we will underestimate radius by factor of $\sim$ $\sqrt{2}$, and underestimate mass by a factor of 2.  However, we consider it unlikely that such an effect is strongly affecting our results, since spots this large are expected to be rather rare.  The usual covering fraction is more like 30\%, which will cause us to underestimate radius by $\sim$ 15\% and mass by $\sim$ 30\%. We discuss the effect of such spots on our results in the main body of the text (\S 4.1.1).  

Our luminosity estimates may also be affected by the presence of spots, either because we infer a \teff that is less than that of the photosphere (by $\sim$ 200K, as discussed above and in Paper I), or because a large fraction of the star appears dark due to a large ultra-cool spot.  However, we emphasize that the luminosity we then infer is not erroneous, but a true indicator of star's energy output.  A star covered by a large cool spot {\it is} less luminous than it would be without the spot (keeping the \teff of the unspotted photosphere constant).  

Finally, we assume that the spot coverage is the same during the photometric and spectroscopic observations, though the two were not obtained simultaneously (separated by $\lesssim$4 years in all cases).  If the spot covering fraction were much larger when the photometry was obtained, then the star will appear redder (i.e., cooler) in the photometric measurements than in the spectroscopic ones.  Consequently, we would overestimate extinction, radius, mass and luminosity.  The opposite occurs if the spot coverage were significantly lower during the photometric observations.  Without knowing a priori the magnitude of temporal variations in spot coverage in our sample, it is impossible to precisely quantify the resultant errors.  

However, as discussed above, spots affect our analysis only if they cover a large fraction ($\gtrsim$ 50\%) of the surface.  If the covering fraction is this large, then its temporal variations might be comparatively small (assuming the spotting is not concentrated in one place). While individual spots may appear and disappear, it is unlikely that a very heavily spotted star will become, over only 4 years, a comparatively unspotted one, or vice versa (stellar cycles are not thought to occur in such young objects).  Thus, assuming the spot coverage is relatively unchanged between the photometric and spectroscopic observations seems fairly reasonable.  If, on the other hand, the spot covering fraction is small ($<<$50\%), its effect on our analysis is minimal.  In this case, it is immaterial whether the spot coverage is the same during the photometric and spectroscopic observations (as long as it is small during both).

\section{Binarity}
We only concern ourselves here with relatively binaries, where potential complications arise from both components appearing in the photometric and spectroscopic measurements.  We examine three cases: equal-brightness binaries, binaries with slightly (500K) cooler secondaries, and binaries with much cooler and fainter secondaries.  Secondaries much cooler (and correspondingly much fainter) than the primary will not affect our derived parameters for the primary. Considering the high binary fraction found by other investigators in low-mass PMS samples in different clusters, binarity may affect our sample as well.  However, given that only relatively close binaries can affect our analysis at all, and moreover that none of our targets are double-lined, the binary phase space that affects us is restricted to single-lined systems, and thus severely curtailed.  In other words, we do not expect a large fraction of our targets to be binaries. We can safely ignore the issue of eclipsing systems, since they are very rare, and quite unlikely to affect our small sample.

For equal-brightness systems, i.e, equal \teff and gravity (assuming coevality), our \teff, gravity and $\av$ estimates will not be affected, since both components contribute similar spectra and photometric fluxes.  However, the observed luminosity will be greater than that from any one component.  In the worst-case scenario, we will overestimate luminosity by a factor of 2, and hence overestimate radius and mass by factors of $\sqrt{2}$ and 2 respectively.

Now consider secondaries cooler by $\sim$ 500K than the primary.  First assume that both have the same radius.  In that case, in analogy to the 500K cooler cool-spot case, we will infer a \teff $\sim$ 200K lower than the primary's.  Our gravity estimate will not be significantly affected even if the secondary's gravity is 0.5 dex smaller (in analogy with a low-gravity cool spot, discussed in Paper I).  Our $\av$ estimates will also be only marginally affected, as in the cool spot case.  However, the observed $J$ magnitude will on average be similar to that from two stars, each cooler than the real primary by $\sim$ 200K (assuming one component is not occluding the other).  The model spectra indicate that, for \teff in the 3000-2500K range, a decrease in \teff by 100K leads to $J$ fainter by $\sim$ 0.15 mag.  The observed $J$-band flux in our case will then differ from the primary's by (2$\times$0.15)-2.5$log$(2), i.e., the system will appear $\sim$ 0.35 mag brighter in $J$ than the primary alone.  We will thus overestimate the primary's luminosity by roughly 40\%, and overestimate its radius and mass by about 20\% and 40\% respectively (very similar conclusions are reached if one uses $\ic$ fluxes instead of $J$ to calculate radius, mass and luminosity).  Of course, if the cooler component has a smaller radius (as is likely for coeval objects), then we would be even less affected by its presence, since its flux contribution would be much lower.  

\section{Extinctions in GG Tau}
One potential cause for concern is the difference in the extinctions we derive for GG Tau Ba and Bb; for Ba we find $\av$=0.29 mag, and for Bb 1.76 mag.  In their analysis of GG Tau, WGRS99 have inferred  $\av$ = 0.55 mag for Ba, and 0.0 for Bb.  Two questions may then be posed.  First, is it plausible that Ba and Bb have visual extinctions differing by 1.5 mag, as we find?  Second, how do we explain the difference between our and WGRS99's estimates?

We address the plausibility issue first.  WGRS99 find an $\av$ of 0.72 mag for GG Tau Aa, 3.20 for Ab, and, as noted above, 0.55 mag for Ba and 0.0 for Bb.  Thus, even their analysis indicates a substantial variation in extinction among the four components of GG Tau.  The large difference in $\av$ that WGRS99 find between Aa and Ab is somewhat surprising, given that they are separated by only 0''.25.  It is possible, as WGRS99 suggest, that this arises from differences in the local distribution of circumstellar material around each star; whether this is actually the case, given that GG Tau A is an almost face-on system, remains an open question.  At any rate, it seems very likely that there is at least a moderate amount of extinction towards both components of GG Tau A, and towards GG Tau Ba as well (as both WGRS99 and we find).  The Taurus star-forming region is also known to be a heavily reddened one.  Under the circumstances, our estimate of $\av$ $\sim$ 1.8 seems unremarkable; it would perhaps be more surprising, given its environs and the observed $\av$ in the other 3 GG Tau members, if GG Tau Bb did not have any extinction  at all.  We further note that the separation between Ba and Bb is 1''.48, much larger than between Aa and Ab; if, as WGRS99 find, even Aa and Ab can differ in $\av$ by $\sim$ 2.5 mag, there is no a priori reason to discard our result of a smaller, 1.5 mag $\av$ difference between the much wider pair Ba and Bb.  Finally, we point out that our errors in extinction are, on average, $\sim$ $\pm$ 0.7 mag\footnote{Our estimated $\av$ error of $\pm$0.7 mag assumes an error of $\sim$0.12 mag in observed $\rc-\ic$ (\S 3.3).  This is accurate for our Upper Sco targets.  For GG Tau Ba and Bb, the errors in observed $\rc$ and $\ic$ quoted by WGRS99 (whose photometry we use for GG Tau) implies an $\av$ error of $\pm$0.4 mag for Ba, and of $\pm$1 mag for Bb (calculated by adding in quadrature the other errors in our analysis, e.g., in determining \teff and gravity; \S 3.2).  This does not create any substantial difference in our conclusions: our $\av$ error for Ba is slightly less that 0.7 mag, and for Bb slightly more, so an average error of 0.7 mag for each is a good estimate.}.  Consequently, our $\av$ results are not incompatible with Ba and Bb actually being more similar to each other in extinction, than our quoted $\av$ values might at first suggest.  We emphasize that we have accounted for such $\av$ uncertainties in our mass, radius and luminosity analysis, so our conclusions regarding the latter quantities are unaffected by our discussion here.
  
The second question is why our $\av$ estimate for Bb differs significantly from that of WGRS99.  The answer lies in the difference in methodology used in the two studies.  Our method has already been discussed in \S 3.1.  Basically, our high-resolution spectral analysis allows us to first derive gravity and \teff independent of extinction, through comparison with synthetic spectra (Paper I); we then adopt the intrinsic colors implied by the synthetic spectra (for the derived \teff and log \gv), and compare to the observed colors, to infer $\av$.  WGRS99, on the other hand, calculate $\av$ on the basis of spectral type considerations.  For GG Tau Bb, they find a spectral type of M7 (slightly different from the more recent, and more accurate estimate of M7.5 by White \& Basri 2002 and Luhman 1999).  To derive $\av$, they then use as a color template the M7 field dwarf VB8 (GJ 644C), which has $\rc-\ic$ = 2.41 \citep{Kirk94}.  The observed $\rc-\ic$ of GG Tau Bb, according to WGRS99, is 2.46.  Using VB8 as a template then implies, through our eqn[1] (\S 3.1) for calculating extinction,  an $\av$ of 0.24 mag for Bb, quite close to the $\av$ = 0.0$\pm$0.24 mag that WGRS99 quote\footnote{WGRS99 use 3 colors ($V-\rc$, $\rc-\ic$, $\ic-J$) to derive $\av$, not $\rc-\ic$ alone.  Consequently, their stated $\av$ $\approx$ 0 $\pm$ 0.24 mag is slightly (but not significantly) different from the $\av$ $\approx$ 0.24 they would have found using $\rc-\ic$ alone.  To facilitate comparison to our $\av$ results (derived from just $\rc-\ic$), we concentrate on $\rc-\ic$ in our present discussion.}.  Now, according to the latest \teff scale for field M dwarfs (Leggett et al. 2000; see discussion in Paper I), an M7 dwarf should have \teff $\lesssim$ 2500K.  Assuming \teff = 2500K, a reasonable gravity of $\sim$ 5.0, and solar-metallicity, the synthetic spectra then predict an $\rc-\ic$ color of $\sim$ 2.3 for VB8 (see Fig. 5). In other words, the predicted $\rc-\ic$ color of VB8, based on spectral type, is already quite close to the observed color; slightly lower \teff, slightly higher gravity and/or metallicity effects are likely to account for any remaining difference.  This line of reasoning shows that, {\it if} we adopted the WGRS99 method of using field dwarf colors as a template to calculate extinction in Bb, we would arrive at a very similar result (i.e., very low extinction) even using our synthetic photometry. 

The real issue, therefore, is whether a field dwarf of the same spectral type accurately represents the intrinsic colors of GG Tau Bb.  If PMS objects are $\sim$ 100-150K hotter than dwarfs of the same type (as suggested by WGRS99 themselves, as well as Luhman 1999), then GG Tau Bb (M7.5) should roughly have \teff $\gtrsim$2500K (consistent with our derived \teff of $\sim$2600K; Paper I)\footnote{WGRS99 find a \teff of $\sim$ 2800K for Bb, which is $\sim$ 300K hotter than an M7 dwarf according to the latest M dwarf \teff scale \citep{Leggett00}.  WGRS99's argument that PMS objects are $\sim$ 100-150K hotter than dwarfs of the same spectral type is in fact based on the older \teff scale for dwarfs by Leggett et al. 1996; the older scale was $\sim$ 200K hotter than the new one.  However, our \teff of $\sim$ 2600K for Bb is indeed only $\sim$ 100K hotter than the new scale.  That is, the difference between WGRS99's value for Bb and the old M dwarf \teff scale is the same as the difference between our value for Bb and new M dwarf scale - $\sim$100K in either case.  These considerations are discussed in detail in Paper I.}.  Fig. 5 shows that, according to the synthetic spectra, PMS objects (log $\gv$ $\lesssim$4.0) with \teff $\gtrsim$ 2500K are bluer by $>$0.2 mag in $\rc-\ic$ than VB8; the latter thus seems to be an unreliable template for GG Tau Bb.  Indeed, the same problem is evident, and in fact exacerbated, if we assign to Bb the \teff that WGRS99 themselves find.  The latter authors compare their observations to the BCAH98 evolutionary tracks to derive \teff $\sim$2800K and age $\sim$1.5 Myr for Bb .  These {\it same} tracks, however, also state that an object with this \teff and age should have an intrinsic $\rc-\ic$ $\sim$2.0 - i.e., much bluer than VB8 (see BCAH98; our synthetic spectra also suggest $\rc-\ic$ $\sim$2.0 for a PMS object at $\sim$ 2800K, see Fig. 5).  Even a field dwarf at 2800K has $\rc-\ic$ $\sim$2.1 (Fig. 5), considerably bluer than VB8.  These considerations imply that by relying on spectral type, WGRS99 have ascribed intrinsic colors to GG Tau Bb that are too red, and thus underestimated its $\av$.  

Under the circumstances, one might ask why a similar discrepancy is {\it not} seen between our and WGRS99's values for $\av$ in GG Tau Ba: we find a visual extinction of 0.29 mag, while WGRS99 find an almost identical (within the errors) value of 0.55 mag\footnote{As in Bb, the WGRS99 $\av$ for Ba is from a 3-color analysis, using template dwarf colors from Kirkpatrick \& McCarthy 1994; $\rc-\ic$ alone would imply $\av$=0.39 in the WGRS99 scheme, even closer to our value.}.  The answer once again lies in spectral typing uncertainties.  WGRS99 derive a spectral type of M5 for Ba, and thus compare its colors to those of M5 dwarfs from the literature.  Specifically, they use M5 dwarf colors from Kirkpatrick \& McCarthy 1994, who quote $\rc-\ic$ = 1.89, calculated by averaging the colors of the two M5 dwarfs Gl 51 and GJ 1057 \footnote{One of these, GL 51, has been analysed by Leggett et al. 2000, and found to have a \teff of 2900K; it is plotted in Fig. 5 (\teff = 2900K, $\rc-\ic$ = 1.92) along with another M5 dwarf analysed by the same authors, GJ 1029 (\teff = 2900K, $\rc-\ic$ = 1.93).}.  However, the spectral type of Ba has been pushed down in the intervening years; Luhman 1999 found M5.5 through a more careful analysis, and the latest estimate is M6, by White \& Basri 2002.  WGRS99 have therefore compared Ba to a field dwarf with an earlier spectral type, and thus bluer in $\rc-\ic$, than appropriate within their scheme for deriving $\av$.  At the same time, as we have noted above in our discussion of GG Tau Bb, dwarfs of a given spectral type seem to be redder in the optical than PMS objects of the same type: higher gravity alone, at a fixed \teff, leads to redder $\rc-\ic$ (Fig. 5); it may also be that PMS objects are systematically somewhat hotter than dwarfs of the same type (Luhman 1999; our results in Paper I generally agree with this view), which exacerbates the color difference.  In general, this causes an underestimation of $\av$ in the WGRS99 scheme (as was the case for Bb).  In the case of Ba, however, WGRS99's use of a dwarf template that is {\it earlier} in type than Ba offsets this effect, producing an $\av$ quite similar to ours.  Note that if WGRS99 had used an M6 field dwarf as template, as dictated by the newest spectral type for Ba, they would indeed have found a substantially lower $\av$ than we do, analogous to the situation for Bb.  For instance, Kirkpatrick \& McCarthy 1994 quote an average $\rc-\ic$ of 2.17 mag for M6 dwarfs.  Using this, WGRS99 would have found no extinction at all.  In fact, strictly speaking, using the M6 dwarf color yields an unphysical negative extinction ($\av$ = -1) for Ba (which has $\rc-\ic$ = 1.97 according to WGRS99).  This is further evidence that dwarf optical colors are intrinsically redder than PMS ones at these spectral types.  A detailed discussion of color differences between dwarf, giant and PMS regimes can be found in Luhman 1999.   

The above discussion clearly illustrates the problems, alluded to in Appendix A, of using spectral type considerations for PMS analysis.  Our methodology, in Paper I and here, avoids these difficulties by making no a priori assumptions about intrinsic PMS colors based on spectral typing: \teff and gravity are first inferred in an extinction-independent fashion, and the extinctions then derived self-consistently, by adopting the intrinsic colors implied by the same synthetic spectra that are used for the \teff and gravity analysis.

\plotone{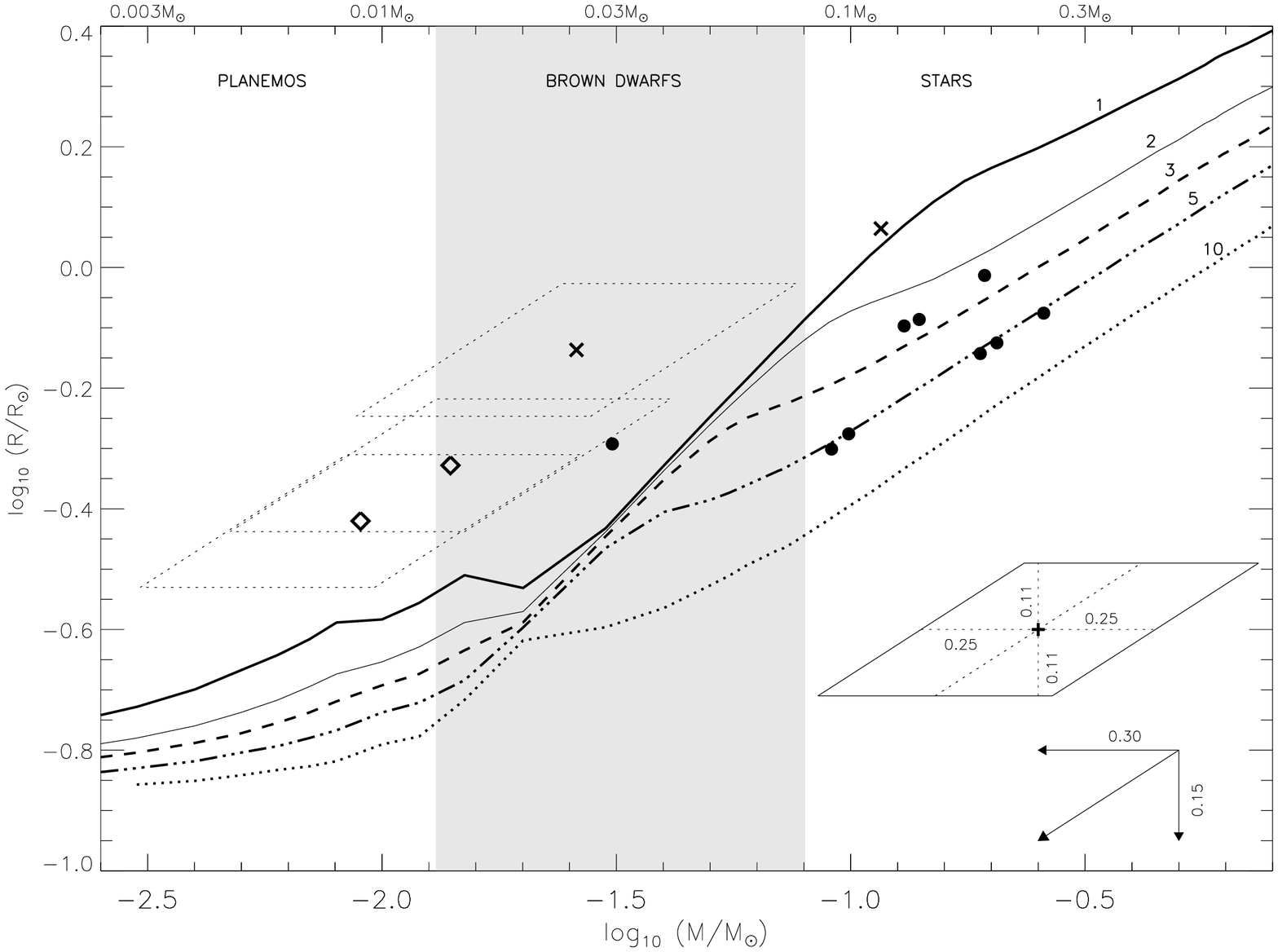}
\figcaption{\label{massrad} Our derived radii and masses (based on $J$ fluxes; see \S3.1), plotted against the Lyon98/00 isochrones (track ages denoted in Myr units).  The central grey region delineates the brown dwarf regime, while the regions to the left and right of it indicate the planemo and stellar regimes respectively.  Diamonds represent the two Upper Sco objects near the planemo boundary (USco 128 \& 130); filled circles indicate all other Upper Sco targets.  The crosses denote GG Tau Ba and Bb (Ba has the larger mass).  Error bars are indicated: the vertical line represents $\pm$0.11 dex errors in radius; the horizontal line represents the error in mass at fixed radius, due to our $\pm$0.25 dex uncertainty in gravity.  At a fixed log \gv, changing radius makes an object move diagonally in the mass-radius plane (since our mass depends on radius), as shown by the diagonal lines superimposed on the error bars.  For the 3 targets with the largest offset from the tracks, the error boundaries are also superimposed on our data points to clearly illustrate their disagreement with the tracks.  The horizontal and vertical arrows at the bottom right show, respectively, the shift in inferred mass and radius that would result if any object were actually an equal-mass binary; the diagonal arrow indicates the combined shift.  Masses $\gtrsim$ 0.03 \msun agree with the Lyon radius predictions for the expected ages, while lower masses have significantly larger radii than predicted.  See \S4.1. }

\plotone{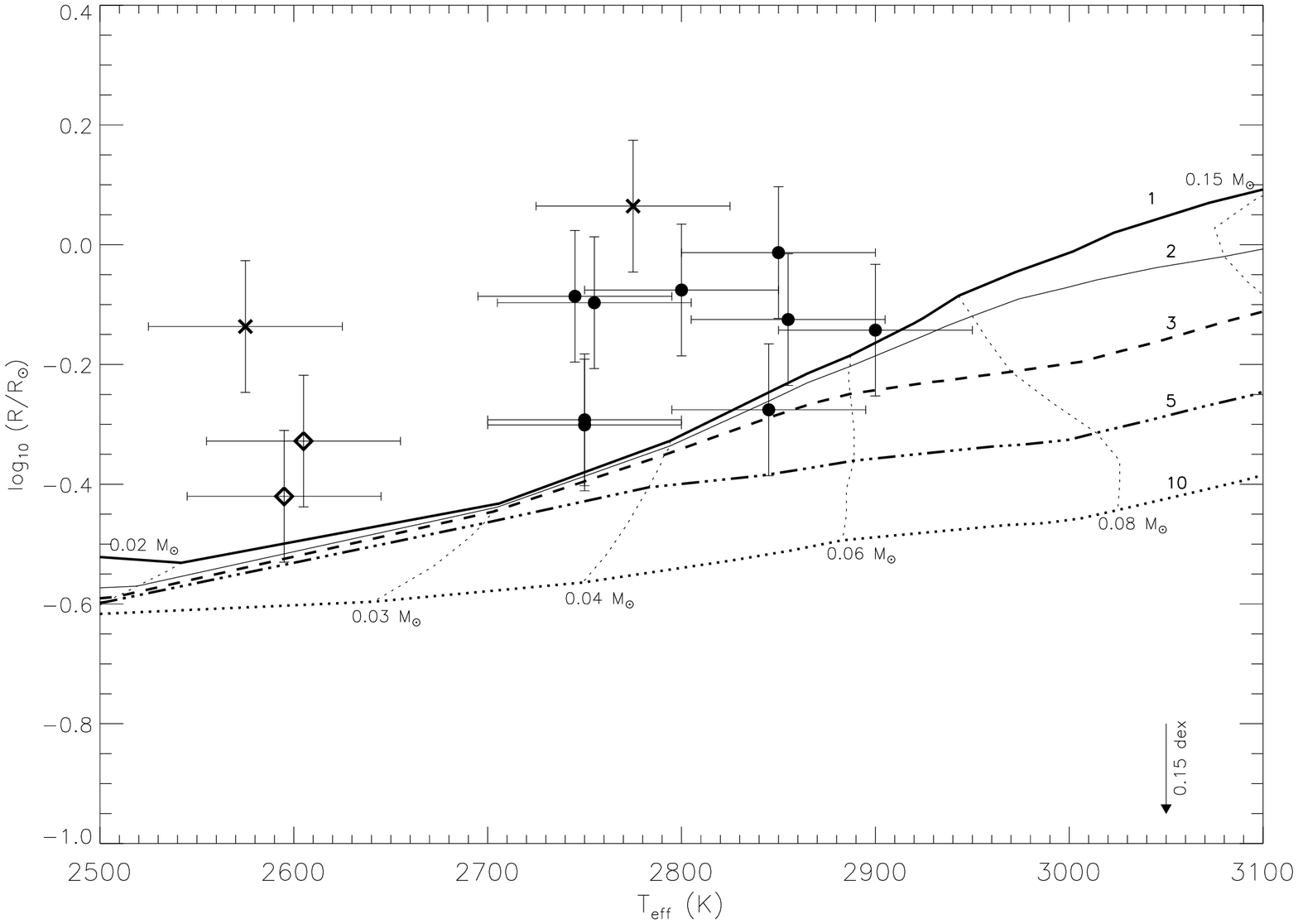}
\figcaption{\label{temprad} Our derived \teff and radii (latter based on $J$ fluxes), versus the Lyon98/00 predictions.  All symbols are the same as in Fig. 1.  The theoretical evolutionary paths for various masses are also plotted (dotted lines); note that these are {\it not} necessarily the masses we derive for our targets.  Error bars are $\pm$ 50K in \teff, and $\pm$ 0.11 dex in radius.  The vertical arrow at the bottom right indicates the shift in inferred radius that would result if any target were an equal mass binary.  While all the objects appear displaced from the Lyon isochrones, the reasons for this vary: the 3 coolest objects (GG Tau Bb and USco 128 \& 130) are offset because their derived radii are much larger than predicted, while the other, hotter targets are offset because their inferred temperatures are much lower than predicted (as apparent by comparing to Figs. 1 and 3; see also \S 4.2). }   

\plotone{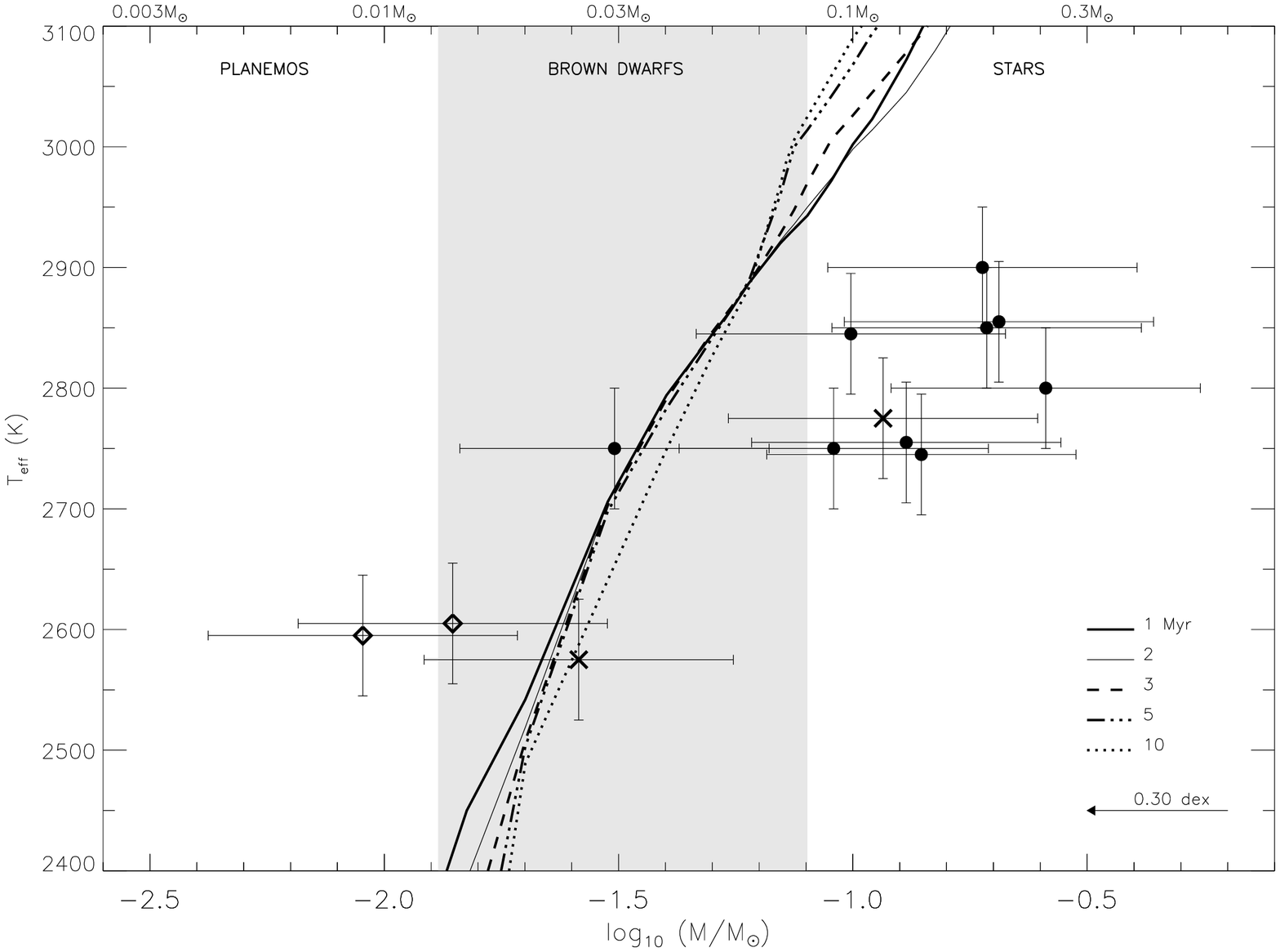}
\figcaption{\label{tempmass} Our derived \teff and masses (latter based on $J$ fluxes), versus the Lyon98/00 predictions.  All symbols are the same as in Fig. 1.  Error bars are $\pm$ 50K in \teff, and $\pm$ 0.33 dex in mass.  The arrow at the bottom right indicates the shift in inferred mass that would result if any object were an equal mass binary.  Our inferred \teff-mass relationship appears shallower than predicted: the two lowest mass targets appear hotter than expected; the two intermediate masses in the brown dwarf regime agree quite well with the Lyon \teff, and the more massive targets all appear much cooler than predicted for their mass.  See \S 4.2.}

\plotone{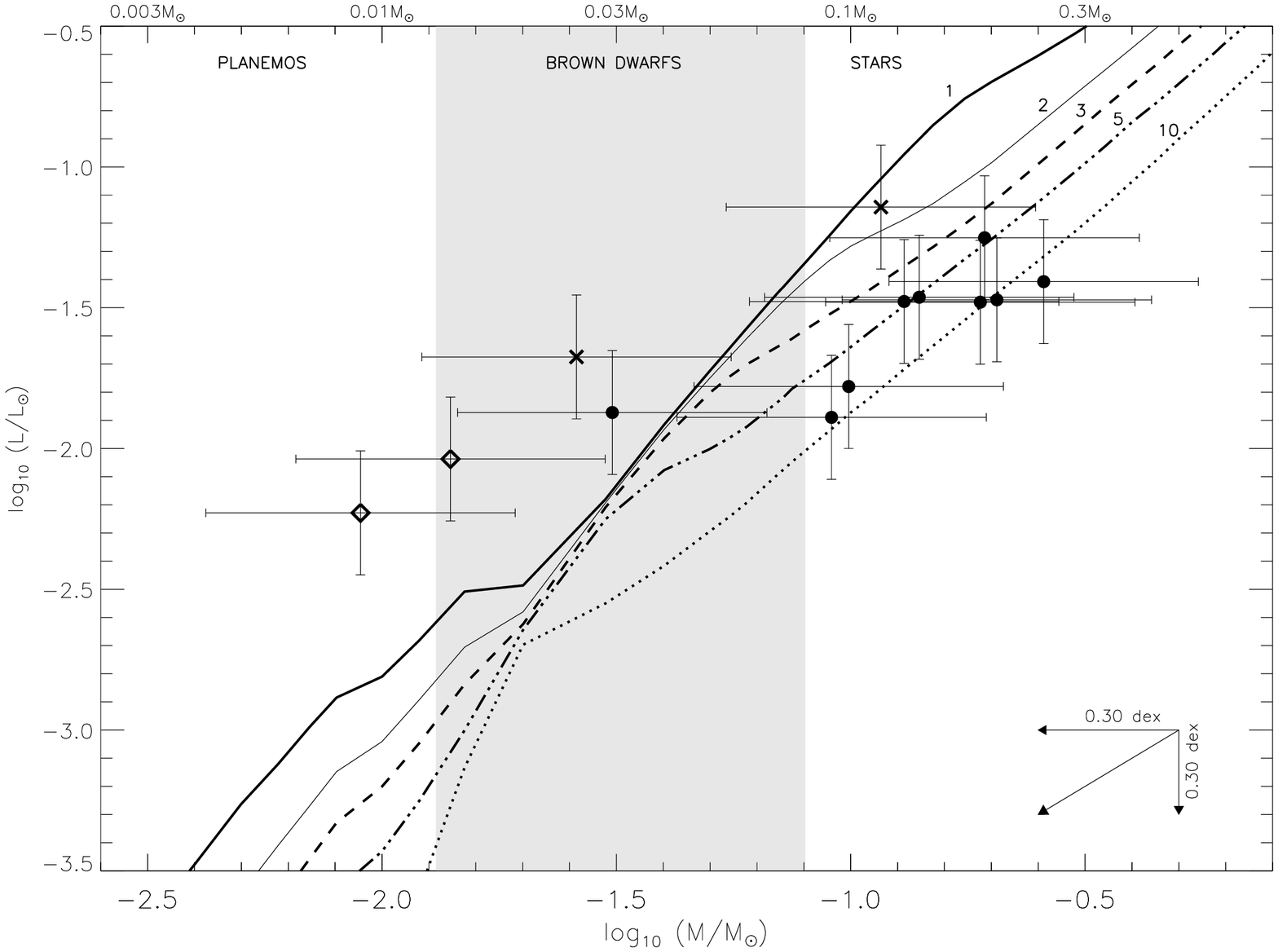}
\figcaption{\label{masslum} Our derived masses and luminosities (based on $J$ fluxes), versus the Lyon98/00 predictions.  All symbols are the same as in Fig. 1.  Errors are $\pm$ 0.33 dex in mass and $\pm$ 0.22 dex in $\lum$.  The horizontal and vertical arrows at the bottom right indicate, respectively, the shift in inferred luminosity and mass that would result if any object were an equal mass binary; the diagonal arrow indicates the combined shift.  See \S 4.3. }

\plotone{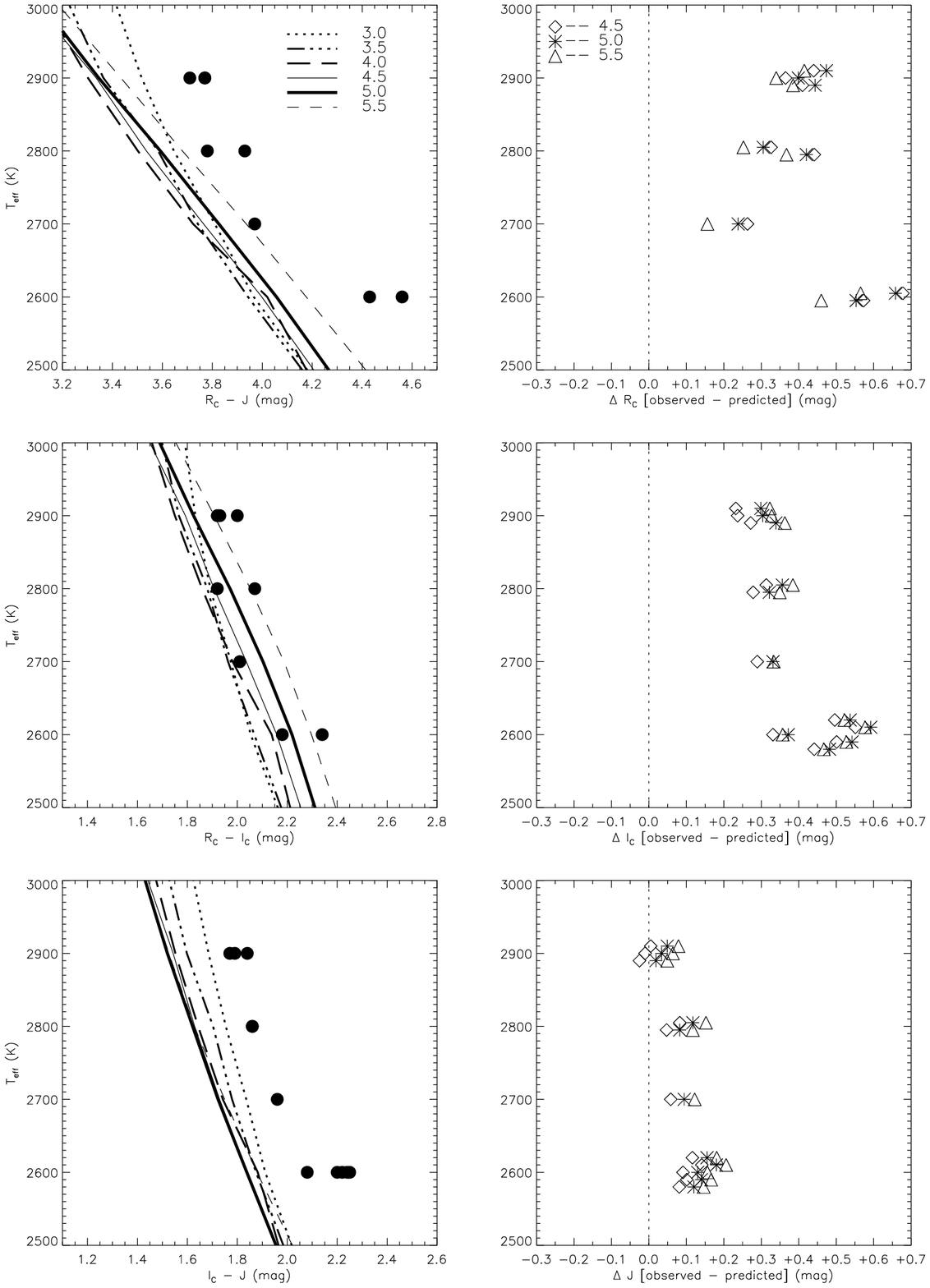}
\figcaption{\label{tempircol} Derived \teff-photometry relationships for field dwarfs, compared to synthetic spectra predictions for various gravities.  Data from Leggett et al. (2000).  The expected gravity of field dwarfs is log \gv $\approx$ 4.5--5.5.  {\it Left column}: \teff-color relationship, for various colors: $\rc-J$ (top), $\rc-\ic$ (middle) and $\ic-J$ (bottom).  The synthetic spectra predictions are too blue in $\rc-J$ and $\ic-J$, but match the observations quite well in $\rc-\ic$.  {\it Right column}: \teff-flux relationship, for various photometric bands: $\rc$ (top), $\ic$ (middle) and $J$ (bottom).  In each panel, we show the observed flux minus the predicted flux (corrected for known distance) at the derived \teff, for three gravities covering the expected range of dwarf gravities (triangle, asterisk and diamond denote log\gv = 4.5, 5.0 and 5.5 respectively).  If the synthetic flux matched the observations exactly for some gravity at a given \teff, the corresponding data point would lie on the vertical dotted line (which corrssponds to zero offset).  We see that the model spectra are overluminous, compared to the observations, in both $\rc$ and $\ic$, but perform quite well (errors $\lesssim$ 0.2 mag) in $J$. See Appendix A. These plots provide the rationale for our choosing $\rc-\ic$ colors to derive extinctions for our targets, and $J$-band photometry to derive radii, masses and luminosities.  }

\clearpage

\begin{deluxetable}{lcccccccccc}
\tablecaption{\label{tab1} Temperatures, Gravities, Photometry \& Extinctions}
\tablewidth{0pt}
\tablehead{
\colhead{name} &
\colhead{SpT\tablenotemark{a}}  &
\colhead{\teff\tablenotemark{b}} &
\colhead{log $\gv$\tablenotemark{b}} &
\colhead{$\rc$\tablenotemark{c}}  &
\colhead{$\ic$\tablenotemark{c}}  &
\colhead{$J$\tablenotemark{d}}  &
\colhead{$\av$\tablenotemark{e}} &
\colhead{$\ai$\tablenotemark{e}} &
\colhead{$\exj$\tablenotemark{e}} &\\
 & &(K) & &(mag) &(mag) &(mag) & (mag) & (mag) & (mag) \\}   
                            
\startdata

USco 55   &M5.5  &2800 &4.00  &16.70  &14.58 &12.46 &1.24 &0.74  &0.35  \\
USco 66   &M6    &2900 &4.00  &17.03  &14.85 &12.91 &2.05 &1.23  &0.58  \\
USco 53   &M5    &2850 &3.75  &16.76  &14.54 &12.29 &1.90 &1.14  &0.54  \\
USco 75   &M6    &2850 &4.00  &17.25  &15.08 &12.78 &1.71 &1.03  &0.48  \\
USco 67   &M5.5  &2750 &3.75  &16.99  &14.87 &12.54 &0.95 &0.57  &0.27  \\
USco 100  &M7    &2750 &3.75  &17.95  &15.62 &12.84 &1.95 &1.17  &0.55  \\
USco 109  &M6    &2750 &4.00  &18.21  &16.06 &13.61 &1.10 &0.66  &0.31  \\
USco 112  &M5.5  &2850 &4.00  &18.22  &16.14 &13.46 &1.29 &0.77  &0.36  \\
USco 104  &M5    &2750 &3.50  &17.76  &15.68 &13.48 &0.76 &0.46  &0.21  \\
USco 130  &M7.5  &2600 &3.25  &19.91  &17.45 &14.20 &1.90 &1.14  &0.54  \\
USco 128  &M7    &2600 &3.25  &19.38  &17.09 &14.40 &1.10 &0.66  &0.31  \\
\\
GGTau Ba  &M6    &2775 &3.375  &15.36  &13.39 &11.48 &0.29 &0.17  &0.08  \\
GGTau Bb  &M7.5  &2575 &3.125  &18.01  &15.55 &13.16 &1.76 &1.06  &0.50  \\

\enddata
\tablenotetext{a}{Spectral Types for Upper Sco objects from AMB00, and for GG Tau Ba and Bb from White \& Basri 2002.}
\tablenotetext{b}{\teff and gravity derived in Paper I; errors $\approx$ $\pm$ 50K and $\pm$ 0.25 dex respectively; derived assuming solar meatallicity ([M/H]=0).}
\tablenotetext{c}{Observed $\rc$ and $\ic$, uncorrected for extinction; taken from AMB00 for Upper Sco objects, and from WGRS99 for GG Tau Ba and Bb.}
\tablenotetext{d}{Observed $J$, uncorrected for extinction; taken from 2MASS for Upper Sco objects, and from WGRS99 for GG Tau Ba and Bb.}
\tablenotetext{e}{Extinctions from comparing the observed $\rc-\ic$ colors to synthetic ones (for the derived \teff and log \gv); error in $\av$ $\approx$ $\pm$ 0.7 mag.  $\ai$ = 0.60 $\av$; error $\approx$ $\pm$ 0.42 mag.  $\exj$ = 0.28 $\av$; error $\approx$ $\pm$ 0.20 mag. }

\end{deluxetable}

\begin{deluxetable}{ccccccccc}
\tablecaption{\label{tab2a} Radii, Masses \& Luminosities}
\tablewidth{0pt}
\tablehead{

\colhead{} & \multicolumn{8}{c}{\hspace{75pt} $\ic$ \hspace{95pt} $J$} \\
\cline{5-5} \cline{8-8} \\ 
\colhead{name} &
\colhead{\teff} &
\colhead{log $\gv$} &
\colhead{$\frac{\mass}{{\rm M}_{\odot}}$} &
\colhead{$\frac{\rad}{{\rm R}_{\odot}}$} &
\colhead{$\frac{\lum}{{\rm L}_{\odot}}$} &
\colhead{$\frac{\mass}{{\rm M}_{\odot}}$} &
\colhead{$\frac{\rad}{{\rm R}_{\odot}}$} &
\colhead{$\frac{\lum}{{\rm L}_{\odot}}$} \\}


\startdata

USco 55  &2800 &4.00  &0.24    &0.81 &0.036   &0.26    &0.84 &0.039 \\
USco 66  &2900 &4.00  &0.24    &0.81 &0.042   &0.19    &0.72 &0.033 \\
USco 53  &2850 &3.75  &0.19    &0.96 &0.055   &0.19    &0.97 &0.056 \\
USco 75  &2850 &4.00  &0.18    &0.70 &0.029   &0.21    &0.75 &0.033 \\
USco 67  &2750 &3.75  &0.10    &0.70 &0.025   &0.14    &0.81 &0.034 \\
USco 100 &2750 &3.75  &0.088   &0.66 &0.022   &0.14    &0.81 &0.034 \\
USco 109 &2750 &4.00  &0.063   &0.41 &0.009   &0.091   &0.50 &0.013 \\
USco 112 &2850 &4.00  &0.053   &0.38 &0.009   &0.099   &0.52 &0.016 \\
USco 104 &2750 &3.50  &0.025   &0.47 &0.011   &0.031   &0.52 &0.014 \\
USco 130 &2600 &3.25  &0.007   &0.33 &0.004   &0.014   &0.47 &0.009 \\
USco 128 &2600 &3.25  &0.006   &0.31 &0.004   &0.009   &0.38 &0.006 \\
\\ 
GGTau Ba &2775 &3.375  &0.11   &1.12 &0.067   &0.12   &1.16 &0.072 \\
GGTau Bb &2575 &3.125  &0.028  &0.75 &0.022   &0.026  &0.73 &0.021 \\

\enddata
\tablecomments{Mass, radius and bolometric luminosity derived in this paper.  The first set of values is derived using $\ic$ fluxes, the second set using $J$.  The internal stochastic errors in each are of order factor of 2 in mass, 30\% in radius, and 65\% in luminosity.  All quantities are derived assuming solar metallicity.  There is a systematic offset between the $\ic$- and $J$-derived values, with the latter generally being somewhat higher.  Objects within a cluster (Upper Sco or Taurus) are listed in order of decreasing $\ic$-band mass; the ordering of objects by $J$-band mass is generally the same.  The few most massive and/or most luminous USco objects may be spectroscopic binaries.  Note that in Figs. 2--5, we have plotted the logarithm of mass, radius and luminosity; this table lists the linear values.}.  

\end{deluxetable}


\clearpage
\begin{thebibliography}{}
\bibitem[Allard, Hauschildt \& Schwenke 2000]{Allard00}
Allard,F., Hauschildt,P.H., Schwenke,D., 2000, \apj, 540, 1005
\bibitem[Ardila, Mart\'\i n \& Basri 2000]{Ardila00}
Ardila,D., Mart\'\i n,E.L., Basri,G., 2000, \aj, 120, 479 [AMB00]
\bibitem[Baraffe et al. 2003]{Baraffe03}
Baraffe,I., Chabrier,G., Barman,T.S., Allard,F., Hauschildt,P.H., 2003, \aap, 406, 1001
\bibitem[Baraffe et al. 2002]{Baraffe02}
Baraffe,I., Chabrier,G., Allard,F., Hauschildt,P.H., 2002, \aap, 382, 563 [BCAH02]
\bibitem[Baraffe et al. 1998]{Baraffe98}
Baraffe,I., Chabrier,G., Allard,F., Hauschildt,P.H., 1998, \aap, 337, 403 [BCAH98]
\bibitem[Basri 2003]{Basri03}
Basri,G., {\it What is a `Planet'?}, 2003, Mercury, vol. 32, 6, p.27
\bibitem[Bate et al. 2002]{Bate02}
Bate,M.R., Bonnell,I.A., Bromm,V., 2002, \mnras, 332, L65
\bibitem[Boss 2001]{Boss01}
Boss,A.P., 2001, \apjl, 551, L167 
\bibitem[Briceno et al. 2002]{Briceno02} 
Briceno,C., Luhman,K., Hartmann,L., Stauffer,J., Kirkpatrick,D., 2002, \apj, 580, 317
\bibitem[Burrows, Sudarsky \& Hubbard 2003]{Burrows03}
Burrows,A., Sudarsky,D., Hubbard,W.B., 2003, \apj, 594, 545
\bibitem[Burrows \& Liebert 1993]{Burrows93}
Burrows,A., Liebert,J., 1993, Rev. Mod. Phys., 65, 301
\bibitem[Carpenter 2001]{Carpenter01}
Carpenter,J.M., 2001, \aj, 121, 2851 
\bibitem[Chabrier et al. 2000]{Chabrier00}
Chabrier,G., Baraffe,I., Allard,F., Hauschildt,P.H., 2000, \apj, 542, 464 [CBAH00]
\bibitem[Charbonneau et al. 2002]{Charbonneau02}
Charbonneau,D., Brown,T.M., Noyes,R.W., Gilliland,R.L., 2002, \apj, 568, 377
\bibitem[Close et al. 2002]{Close02}
Close,L.M., Potter,D., Brandner,W., Lloyd-Hart,M., Liebert,J., Burrows,A., Siegler,N., 2002, \apj, 566, 1095
\bibitem[D'Alessio et al. 1999]{dalessio99}
D'Alessio,P., Calvet,N., Hartmann,L., Lizano,S., Cant\'{o},J., 1999, \apj, 527, 893
\bibitem[D'Alessio, Calvet \& Hartmann 2001]{dalessio01}
D'Alessio,P., Calvet,N. \& Hartmann,L., 2001, \apj, 553, 321
\bibitem[Fern\'{a}ndez \& Comer\'{o}n 2001]{Fernandez01}
Fern\'{a}ndez,M. \& Comer\'{o}n,F., 2001, \aap, 380, 264
\bibitem[Guilloteau, Dutrey \& Simon 1999]{Guilloteau99}
Guilloteau,S., Dutrey,A., Simon,M., 1999, \aap, 348, 570 
\bibitem[Gullbring et al. 1998]{Gullbring98}
Gullbring,E., Hartmann,L.W., Brice\~{n}o,C., Calvet,N., 1998, \apj, 492, 323
\bibitem[Hartigan, Strom \& Strom 1994]{Hartigan94}
Hartigan,P., Strom,K.M., Strom,S.E., 1994, \apj, 427, 961
\bibitem[Hartmann 2003]{Hartmann03}
Hartmann,L., 2003, \apj, 585, 398
\bibitem[Herbig 1998]{Herbig98}
Herbig,G.H., 1998, \apj, 497, 736
\bibitem[Jayawardhana, Mohanty \& Basri 2002]{JMB02}
Jayawardhana,R., Mohanty,S., Basri,G. 2002, \apjl, 578, L141
\bibitem[Jayawardhana, Mohanty \& Basri 2003]{JMB03}
Jayawardhana,R., Mohanty,S., Basri,G. 2003, \apj, 592, 282 
\bibitem[Jayawardhana et al. 2003]{Jaya03}
Jayawardhana,R., Ardila,D., Stelzer,B., Haisch,K.E.,Jr., 2003, \aj, 126, 1515 
\bibitem[Kenyon, Dobrzycka \& Hartmann 1994]{Kenyon94}
Kenyon,S.J., Dobrzycka,D., Hartmann,L., 1994, \aj, 108, 1872
\bibitem[Kirkpatrick et al. 1999]{Kirk99}
Kirkpatrick,J.D. et al., 1999, \apj, 519, 802
\bibitem[Kirkpatrick, Henry \& Irwin 1997]{Kirk97}
Kirkpatrick,J.D., Henry,T.J., Irwin,M.J., 1997, \aj, 113, 1421
\bibitem[Kirkpatrick et al. 1995]{Kirk95}
Kirkpatrick,J.D., Henry,T.J., Simons,D.A., 1995, \aj, 109, 797
\bibitem[Kirkpatrick \& McCarthy 1994]{Kirk94}
Kirkpatrick,J.D., McCarthy,D.W.,Jr., 1994, \aj, 107, 333
\bibitem[Kirkpatrick et al. 1991]{Kirk91}
Kirkpatrick,J.D., Henry,T.J., McCarthy,D.W., 1991, \apjs, 77, 417
\bibitem[Klein et al. 2003]{Klein03}
Klein,R., Apai,D., Pascucci,I., Henning,Th., Waters,L.B.F.M., \apjl, 593, L57 
\bibitem[Lane et al. 2001]{Lane01}
Lane,B.F., Zapatero Osorio,M.R., Britton,M.C., Mart\'\i n,E.L., Kulkarni,S.R., 2001, \apj, 560, 390
\bibitem[Leggett et al. 2001]{Leggett01}
Leggett,S.K., Allard,F., Geballe,T.R., Hauschildt,P.H., Schweitzer,A., 2001, \apj, 548, 908
\bibitem[Leggett et al. 2000]{Leggett00}
Leggett,S.K., Allard,F., Dahn,C., Hauschildt,P.H., Kerr,T.H., Rayner,J., 2000, \apj, 535, 965
\bibitem[Leggett et al. 1996]{Leggett96}
Leggett,S.K., Allard,F., Berriman,G., Dahn,C.C., Hauschildt,P.H., 1996, \apjs, 104, 117
\bibitem[Leggett 1992]{Leggett92}
Leggett,S.K., 1992, \apjs, 82, 351
\bibitem[Lucas \& Roche 2000]{Lucas00}
Lucas,P.W., Roche, P.F. 2000, \mnras, 314, 858
\bibitem[Luhman 1999]{Luhman99}
Luhman,K.L., 1999, \apj, 525, 466
\bibitem[Mohanty, Jayawardhana \& Navascu\'{e}s (2003)]{Mohanty03} 
Mohanty,S., Jayawardhana,R., Barrado y Navascu\'{e}s, D., 2003, \apjl, 593, L109  
\bibitem[Mohanty et al. 2003a]{Mohanty03a} 
Mohanty,S., Basri,G., Jayawardhana,R., Allard,F., Hauschildt,P., Ardila,D., 2003, \apj, submitted [Paper I]
\bibitem[Montalban et al. 2003]{Montalban03} 
Montalban,J., D'Antona,F., Kupka,F., Heiter,U., 2003, \aap, submitted
\bibitem[Najita et al. 2000]{Najita00}
Najita,J.R., Tiede,G.P., Carr,J.S. 2000, \apj, 541, 977
\bibitem[Navascu\'{e}s, Mohanty \& Jayawardhana (2003)]{Navascues03}
Barrado y Navascu\'{e}s,D., Mohanty,S. \& Jayawardhana,R., 2003, \apj, accepted
\bibitem[Padgett et al. 1999]{Padgett99}
Padgett,D., Brandner,W., Stapelfeldt,K., Strom,S., Terebey,S., Koerner,D., 1999, \aj, 117, 1490
\bibitem[Padoan \& Nordlund 2002]{Padoan02}
Padoan,P. \& Nordlund,\AA., 2002, \apj, 576, 870
\bibitem[Potter et al. 2002]{Potter02}
Potter,D., Mart\'\i n,E.L., Cushing,M.C., Baudoz,P., Brandner,W., Guyon,O., Neuh\"{a}user,R., 2002, \apjl, 567, L133
\bibitem[Preibisch et al. 2002]{Preb02}
Preibisch,T., Brown,A.G.A., Bridges,T., Guenther,E., Zinnecker,H., 2002, \aj, 124, 404
\bibitem[Reipurth \& Clarke 2001]{Reipurth01} 
Reipurth,B. \& Clarke,C., \aj, 122, 432.
\bibitem[Ribas 2003]{ribas03}
Ribas,I., 2003, \aap, 398, 239
\bibitem[Schlegel, Finkbeiner \& Davis 1998]{Schlegel98}
Schlegel,D.J., Finkbeiner,D.P., Davis,M., 1998, \apj, 500, 525
\bibitem[Stassun et al. 2003]{Stassun03}
Stassun,K.G., Mathieu,R.D., Vaz,L.P.R., Stroud,N., Vrba,F.J., 2003, \apj, accepted
\bibitem[Torres \& Ribas 2002]{Torres02}
Torres,G., \& Ribas,I., 2002, \apj, 567, 1140
\bibitem[White \& Basri 2002]{White02}
White,R.J. \& Basri,G., 2002, \apj, accepted
\bibitem[White et al. 1999]{White99}
White,R.J., Ghez,A.M., Reid,I.N., Schultz,G., 1999, \apj, 520, 811 [WGRS99]
\bibitem[Zapatero et al. 2000]{Zapa00}
Zapatero Osorio,M.R., B\'ejar,V.J.S., Mart\'\i n,E.L., Rebolo,R., Barrado y Navascu\'es,D., Bailer-Jones, C.A.L., Mundt,R. 2000, Science, 290, 103

\end{thebibliography}
\end{document}